\documentclass[lettersize,journal]{IEEEtran}
\usepackage{amsmath,amsfonts}
\usepackage{algorithmic}
\usepackage{algorithm}
\usepackage{array}
\usepackage[caption=false,font=footnotesize,labelfont=sf,textfont=sf]{subfig}
\usepackage{textcomp}
\usepackage{stfloats}
\usepackage{url}
\usepackage{verbatim}
\usepackage{graphicx}
\usepackage{cite}
\usepackage{xcolor}
\usepackage{multirow}
\usepackage{makecell}
\usepackage{bbding}
\usepackage{tabularx}

\begin{document}

\title{B-CANF: Adaptive B-frame Coding with Conditional Augmented Normalizing Flows}

\author{\textcolor{black}{Mu-Jung Chen, Yi-Hsin Chen, and Wen-Hsiao Peng,~\IEEEmembership{Senior Member,~IEEE}}\thanks{\textcolor{black}{All authors are affiliated with the Department of Computer Science, National Yang Ming Chiao Tung University, Hsinchu, Taiwan (e-mail: wpeng@cs.nctu.edu.tw).}}
\thanks{This work is supported by National Science and Technology Council, Taiwan (NSTC 111-2634-F-A49-010-, MOST 110-2221-E-A49-065-MY3), MediaTek, and National Center for High-performance Computing.}
        % <-this % stops a space
}

% \markboth{IEEE TRANSACTIONS ON CIRCUITS AND SYSTEMS FOR VIDEO TECHNOLOGY


\IEEEpubid{\begin{minipage}{\textwidth}\ \\[12pt] \centering
  Copyright © 2023 IEEE. Personal use of this material is permitted. \\However, permission to use this material for any other purposes must be obtained from the IEEE by sending an email to pubs-permissions@ieee.org.
\end{minipage}} 

\maketitle

\begin{abstract}

Over the past few years, learning-based video compression has become an active research area. However, most works focus on P-frame coding. Learned B-frame coding is under-explored and more challenging. This work introduces a novel B-frame coding framework, termed B-CANF, that exploits conditional augmented normalizing flows for B-frame coding. B-CANF additionally features two novel elements: frame-type adaptive coding and B*-frames. Our frame-type adaptive coding learns better bit allocation for hierarchical B-frame coding by dynamically adapting the feature distributions according to the B-frame type. Our B*-frames allow greater flexibility in specifying the group-of-pictures (GOP) structure by reusing the B-frame codec to mimic P-frame coding, without the need for an additional, separate P-frame codec. On commonly used datasets, B-CANF achieves the state-of-the-art compression performance as compared to the other learned B-frame codecs and shows comparable BD-rate results to HM-16.23 under the random access configuration in terms of PSNR. When evaluated on different GOP structures, our B*-frames achieve similar performance to the additional use of a separate P-frame codec.

\end{abstract}


\section{Introduction}

\IEEEPARstart{T}{he} great success of learned image compression has spurred a new wave of research and development for learned video compression. Most existing methods~\cite{dvclu,dvcpro,ssf,mlvc,rafc,elfvc,nvc} address low-delay P-frame coding based on hybrid-based coding architecture, which comprises uni-directional temporal prediction followed by deep residual coding. Ladune~\emph{et al.}~\cite{mmsp} recently show that residual coding is sub-optimal from the information-theoretic perspective, proposing to condition the variational autoencoder (VAE)-based deep compression on motion-compensated frames~\cite{mmsp,iclrw} with the aim of reaching the lower conditional entropy.

In contrast to the rapid progress in learned P-frame coding, learned B-frame coding, which allows bidirectional referencing for higher coding efficiency, is an under-explored topic. The coding of B-frames usually involves frame interpolation with explicit motion modeling and coding. Residual or conditional coding may then follow to achieve inter-frame prediction. Table~\ref{tab:related} dissects some notable B-frame coding schemes in terms of their strategies for motion and inter-frame coding, to give an overview of recent developments in this area. As shown, most of the existing approaches~\cite{Djelouah, hlvc, BEPIC, murat_lhbdc} adopt residual-based inter-frame coding, while many still use intra motion coding (e.g. encoding optical flow maps as individual images). Notably, the first B-frame coding scheme~\cite{Wu_2018} already adopts conditional inter-frame coding, although its motion coding is still an intra-based approach. Built on top of~\cite{mmsp}, the more recent VAE-based conditional coding framework~\cite{iclrw} features both conditional motion and inter-frame coding. %It, however, inherits the same bottleneck issue from \cite{mmsp}. As such, 
However, its compression performance is inferior to the two state-of-the-art methods~\cite{BEPIC, murat_lhbdc}, both adopt residual-based inter-frame coding. We argue that the full potential of conditional coding is yet to be seen. How it is implemented can make a big difference in coding performance.

Inspired by our previous work~\cite{CANF-VC}, we propose a novel learned B-frame coding framework, termed B-CANF, that exploits conditional augmented normalizing flows (CANF) for B-frame coding. We choose CANF as the coding backbone because of its \textit{generality} and \textit{reversibility}. It is shown in~\cite{CANF-VC} that CANF is able to achieve greater expressiveness by stacking multiple conditional VAE encoders and decoders.

\textcolor{black}{B-CANF extends~\cite{CANF-VC} and differs from most prior works (e.g.~\cite{dvclu,dvcpro,ssf,mlvc,rafc,elfvc,nvc,lvc,compundspr,resolAda}) in several significant ways: 
\begin{itemize}
    \item B-CANF addresses primarily B-frame coding rather than P-frame coding. Similar to~\cite{CANF-VC}, it adopts conditional motion and inter-frame coding. However, the way in which the motion and inter-frame predictors are formulated has been adapted to B-frame coding. 
    \item B-CANF newly introduces frame-type adaptive coding that learns better bit allocation for hierarchical B-frame coding. %This is not seen in any learned B-frame coding framework. 
    \item B-CANF newly proposes a special type of B-frame, called B*-frame, to mimic P-frame coding. This tool feature allows greater flexibility in specifying the group-of-pictures (GOP) structure, without the need for an additional, separate P-frame codec.
\end{itemize}}

%\IEEEpubidadjcol
The non-trivial application of these novel elements–-namely, CANF, frame-adaptive coding, and B*-frame–-to B-frame coding achieves higher compression performance.
To our best knowledge, B-CANF is also the first learned B-frame coding scheme that shows comparable PSNR results to HM-16.23 under the random access configuration with intra-period 32 and GOP size 16. Furthermore, this work presents a comprehensive study on B-frame coding, the results of which are not reported in our previous work~\cite{CANF-VC} for P-frame coding. These facts distinct this work from~\cite{CANF-VC} and highlight its contributions to advancing B-frame coding.
\IEEEpubidadjcol

It is to be noted that this work is an expanded version of our conference publication~\cite{ISCAS23}, which adopts CANF for B-frame coding but focuses specifically on YUV 4:2:0 content, as opposed to RGB content in this work. Moreover, this work includes extensive ablation studies and complexity analyses, which are not covered in \cite{ISCAS23}.
%\IEEEpubidadjcol

\textcolor{black}{The remainder of this paper is organized as follows: Section~\ref{sec:related} reviews learned video compression and the basics of ANF-based image compression (ANFIC~\cite{anfic}). Section~\ref{sec:method} elaborates the design of B-CANF. Section~\ref{sec:experiment} compares B-CANF with the state-of-the-art methods and presents ablation experimental results. Finally, we provide concluding remarks in Section~\ref{sec:conclusion}.}

\vspace{-0.3cm}

\section{Related Work}
\label{sec:related}

\begin{table*}[t]
\centering
\setlength{\tabcolsep}{2.9pt}

\caption{Dissection of some notable B-frame coding schemes.} 
\vspace{\baselineskip}

\begin{tabular}{ccccc}
\Xhline{2\arrayrulewidth}
       & \makecell[c]{\textcolor{black}{Publication}} & \makecell[c]{Motion Coding \\ for B-frames}     & \makecell[c]{Inter-frame Coding \\ for B-frames}  &   \makecell[c]{Modules with  \\ Shared Param.}  \\
\Xhline{2\arrayrulewidth}
\cite{Wu_2018}            & ECCV'18        & VAE-based Intra                 & Cond. VAE-based                & No        \\
\hline
\cite{Djelouah}           & ICCV'19        & VAE-based Intra                 & VAE-based Residual             & I, B      \\
\hline
HLVC~\cite{hlvc}          & CVPR'20        & VAE-based Intra \& DIRECT       & VAE-based Residual             & No        \\
\hline
\cite{iclrw}              & ICLRW'21       & One-stage, Cond. VAE-based      & Cond. VAE-based                & I, P, B   \\
\hline
B-EPIC~\cite{BEPIC}       & ICCV'21        & VAE-based Intra                 & VAE-based Residual             & B, P      \\
\hline
LHBDC~\cite{murat_lhbdc}  & TIP'22         & VAE-based Residual              & VAE-based Residual             & No        \\
\hline
B-CANF (Ours)             & -              & Cond. ANF-based                 & Cond. ANF-based                & B, B*     \\             
\Xhline{2\arrayrulewidth}
\end{tabular}
\label{tab:related}
\end{table*}
\vspace{\baselineskip}

\subsection{Learned P-frame Coding}
\label{related:video_compression}
End-to-end learned video compression has recently attracted lots of attention. Most prior works~\cite{dvclu,ssf,lvc,nvc,fvc,mlvc,hlvc,rafc,elfvc,rlvc,accv,alphavc,c2fm} focus on low-delay, P-frame coding, where a target frame is coded based on information propagated from the past decoded frames. In common, they share a similar temporal predictive coding architecture to the conventional codecs~\cite{hevc,vvc}. As such, improving temporal prediction is one of the central research themes. In this aspect, there have been schemes such as motion-compensation networks~\cite{dvclu}, scale-space warping~\cite{ssf}, one-stage motion estimation~\cite{nvc}, feature-domain warping~\cite{fvc}, multi-hypothesis prediction~\cite{mlvc}, compound spatiotemporal representation~\cite{compundspr}, and recurrent autoencoding~\cite{rlvc,accv}. Because temporal prediction often relies on explicit motion modeling, some research efforts are dedicated to reducing motion overhead by predictive motion coding~\cite{mlvc}, incremental flow map coding~\cite{elfvc}, resolution-adaptive flow coding~\cite{rafc,c2fm} and \textcolor{black}{coarse-to-fine motion coding~\cite{c2fm}}. \textcolor{black}{Other notable P-frame coding techniques include the temporal prior~\cite{rlvc,nvc} that leverages a recurrent neural network to propagate causal, temporal information for entropy coding, adaptive residual skip coding~\cite{alphavc,c2fm}, and transformer-based autoencoders~\cite{qual_ttcoding}.}   

More recently, conditional coding~\cite{mmsp,dcvc,tcm, CANF-VC} achieves a breakthrough in inter-frame coding. It conditions the inter-frame autoencoder on the motion-compensated frame in forming a non-linear prediction of the target frame, as opposed to subtracting the motion-compensated frame from the target frame for residual coding, which is shown to be sub-optimal from the information-theoretic perspective~\cite{mmsp}.
\textcolor{black}{Lately, Mentzer~\emph{et al.}~\cite{google_ttcoding} take an interesting approach to conditional coding, utilizing transformers~\cite{transformer} to model the dependencies between the latents of video frames for entropy coding. It has the striking feature of not requiring motion coding and warping operations. Although its initial idea dates back to~\cite{Wu_2018}, research on conditional coding remains very active. }

\subsection{Learned B-frame Coding}

In comparison with learned P-frame coding, learned B-frame coding~\cite{Wu_2018, Djelouah, hlvc, iclrw, BEPIC, murat_lhbdc}, which is the main focus of this paper and targets higher coding efficiency by allowing temporal prediction from both the future and past decoded frames, is relatively under-explored. Its process normally involves frame interpolation with explicit motion modeling and coding, followed by inter-frame residual coding~\cite{Djelouah, hlvc, BEPIC, murat_lhbdc} or conditional coding~\cite{Wu_2018,iclrw}. 
Due to the needs for bi-directional temporal prediction, B-frame coding usually incurs more motion overhead than P-frame coding. VAE-based intra motion coding~\cite{Wu_2018,Djelouah,hlvc}, which encodes (bi-directional) optical flow maps as individual images, is very common. To minimize motion overhead, several strategies are proposed. Yang~\emph{et al.}~\cite{hlvc} reuse the motion of a future or a past reference frame in a way similar to the DIRECT mode in~\cite{h264}. Yılmaz~\emph{et al.}~\cite{murat_lhbdc} adopt motion residual coding, with the predicted flow maps derived from the two reference frames. Ladune~\emph{et al.}~\cite{iclrw} introduce one-stage, conditional motion coding, where the flow maps are estimated directly from and coded conditionally on the two reference frames without the use of optical flow estimation networks. \textcolor{black}{Pourreza~\emph{et al.}~\cite{BEPIC} reuse the P-frame codec for B-frame coding, sending for each B-frame only one flow map that characterizes the motion between the target B-frame and its predicted frame interpolated by a pre-trained network.} Efforts have also been made to share networks for coding different types of frames, e.g. I-frame, P-frame and B-frame~\cite{iclrw,BEPIC}. 

\textcolor{black}{Table~\ref{tab:related} contrasts the differences between our B-CANF and some notable learned B-frame coding schemes. Extending from our previous work~\cite{CANF-VC} for P-frame coding, this paper introduces a conditional coding framework that exploits CANF for B-frame coding. As opposed to the other conditional coding schemes~\cite{Wu_2018,iclrw}, our method applies conditional coding to both motion and inter-frame coding. Moreover, it features frame-type adaptive coding, which adapts the coding process to the B-frame type. 
We also introduce a special type of B-frame, called B*-frame, which mimics P-frame coding by reusing the B-frame codec.}

\begin{figure}[!t]
    \centering
    \includegraphics[width=0.85\linewidth]{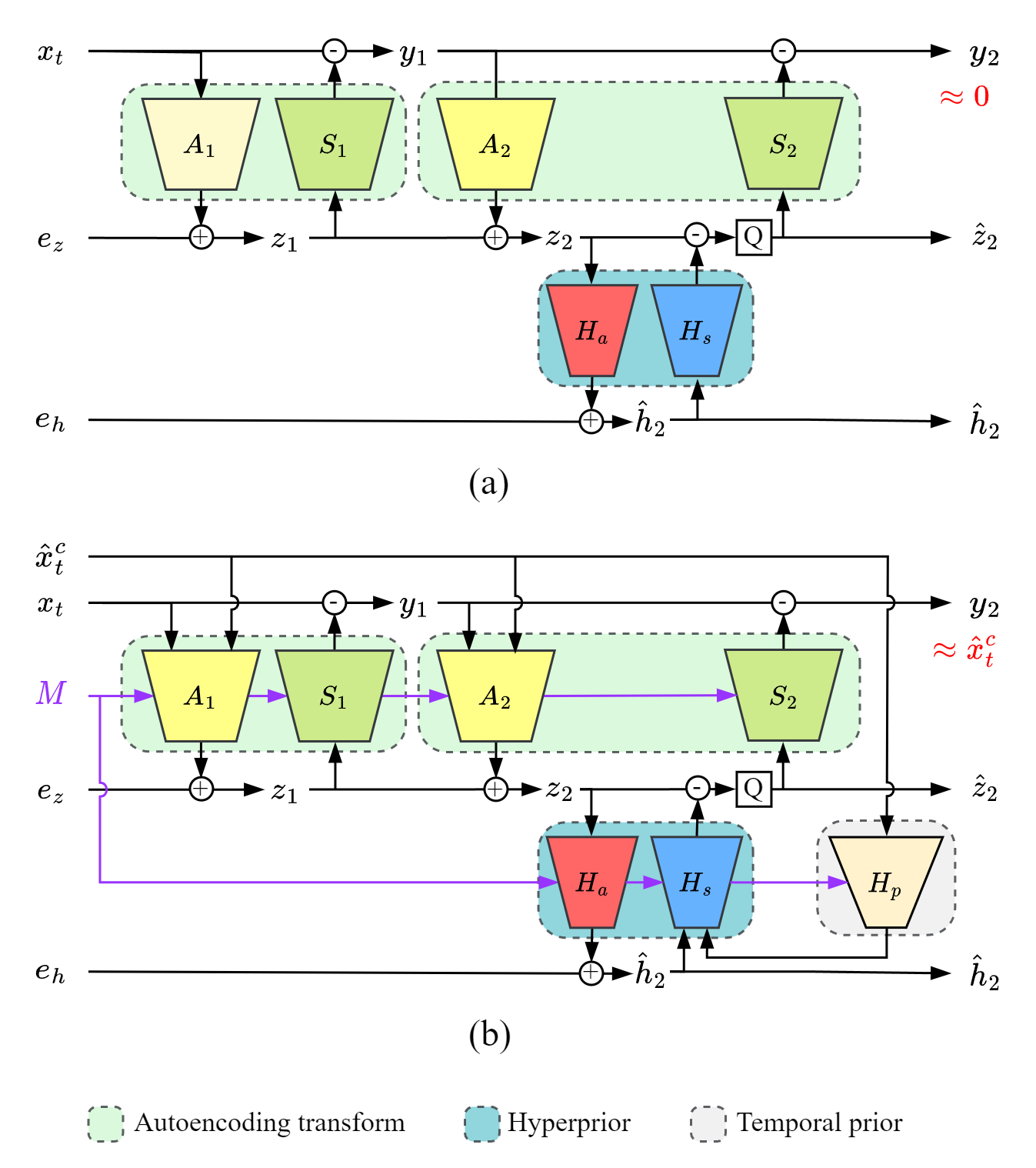}
    \caption{Illustration of (a) the \textit{unconditional} ANF for image coding in ANFIC~\cite{anfic} and (b) the proposed \textit{conditional} ANF for B-frame coding, where \textit{M} signals the B-frame type for frame-type adaptive coding.}
    \label{fig:anfs}
\end{figure}

\subsection{Augmented Normalizing Flows for Image Compression}
B-CANF and our previous work~\cite{CANF-VC} are a sequel to the augmented normalized flow (ANF)-based image compression, also known as ANFIC~\cite{anfic}, which adopts ANF~\cite{anf} as the compression backbone. To ease the understanding of our B-CANF, this section reviews the basics of ANFIC~\cite{anfic}.
Fig.~\ref{fig:anfs}(a) illustrates its coding architecture. Shown on the left are the three inputs of ANFIC~\cite{anfic}: the image to be encoded $x$ and the two augmented inputs $e_z, e_h$; accordingly, the compressed outputs are a nearly zero image $y_2 \approx 0$, the quantized latent code $\hat{z}_2$, and the hyperprior $\hat{h}_2$. 

\textbf{Encoding}: The encoding process proceeds from left to right. The input pair $(x,e_z)$ with $e_z=0$\footnote{ANF~\cite{anf} is initially designed to be a generative model, where the augmented noises are meant to induce a complex marginal on the input $x_t$ and the entropy rate of the latents is not a concern. In ANFIC~\cite{anfic}, the latents $z_2$ must be quantized and compressed to achieve low entropy. Injecting noise at $e_z$ will increase the entropy of $z_2$. ANFIC~\cite{anfic} shows that having $e_z = 0$ can still achieve good compression performance.} 
is transformed into the output pair $(y_2,\hat{z}_2)$, where $y_2$ is regularized to approximate a zero image, by \textcolor{black}{two autoencoding transforms $(A_1, S_1)$ and $(A_2, S_2)$ arranged in the form of additive coupling layers. $A_1, A_2$ are VAE-based analysis transforms, while $S_1, S_2$ are VAE-based synthesis transforms. In symbols, the intermediate states $y_1,z_1$ after the first autoencoding transform $(A_1, S_1)$ are given by}
\color{black}
\begin{align}
    & A_1(x_t,e_z)=(x_t, e_z + \mu_{A_1}(x_t))=(x_t,z_1), %\nonumber 
    \\
    & S_1(x_t,z_1)=(x_t-\mu_{S_1}(z_1), z_1)=(y_1,z_1), %\nonumber 
\end{align}
\color{black}
\textcolor{black}{where $\mu_{A_1},\mu_{S_1}$ are two neural networks that output element-wise additive transform parameters. The second autoencoding transform $(A_2, S_2)$ that converts $y_1,z_1$ into $y_2,\hat{z}_2$ follows a similar definition, except that $z_2$ is additionally predicted from the hyperprior and quantized as $\hat{z}_2$. Here $H_a, H_s$ serve as the hyperprior autoencoder~\cite{googleiclr18}, which encodes the latent code $z_2$ into $\hat{h}_2$ for entropy coding. The reader is referred to \cite{anfic} for network details.} 
\textcolor{black}{By doing so, ANFIC transforms the information from $x_t$ into the augmented space $\hat{z}_2, \hat{h}_2$, which have much lower resolution than the original image and allow for more efficient coding.}

\textbf{Decoding}: The decoding proceeds in reverse order by decoding first the hyperprior $\hat{h}_2$, followed by the decoding of $\hat{z}_2$ to update $y_2$ (initialized as $0$) successively from right to left. During training, $e_h \sim \mathcal{U}(-0.5, 0.5)$ simulates the additive quantization noise of the hyperprior. Remarkably, the conventional VAE-based compression is a special case of ANFIC by retaining only the autoencoding transform $(A_2,S_2)$. In this sense, ANFIC is able to achieve superior expressiveness to VAE-based compression systems by cascading more autoencoding transforms.

\textcolor{black}{Table~\ref{tab:anf_compare} summarizes the differences of ANFIC~\cite{anfic}, our previous work CANF-VC~\cite{CANF-VC}, and B-CANF. Both CANF-VC~\cite{CANF-VC} and B-CANF adopt CANF as the coding backbone, whereas ANFIC uses ANF. CANF differs from ANF in that it learns \textit{conditional} distributions of video frames rather than \textit{unconditional} distributions of images (Section~\ref{method:CANF}). Moreover, B-CANF distinguishes from CANF-VC~\cite{CANF-VC} by the support of B-frames and B*-frames (Section~\ref{sec:B_frame_extension}); additionally, it incorporates frame-type adaptive coding (Section \ref{sec:FrameAdaptCoding}). Extensive experimental results and ablation studies on B-frame coding reported in this paper are not seen in~\cite{CANF-VC}.}

\begin{table}[t]
    \color{black}
    \caption{Comparison of B-CANF, ANFIC~\cite{anfic}, and CANF-VC~\cite{CANF-VC}.}
    \centering
    \setlength{\tabcolsep}{2.9pt}
    \begin{tabular}{ccccc}
    \hline
    \Xhline{2\arrayrulewidth}   & Codec Type & \makecell[c]{Frame-type \\ Support} & \makecell[c]{Frame-type\\ Adaptive Coding} \\
    \hline
    ANFIC~\cite{anfic} & ANF-based  &  I  & N\\
    CANF-VC~\cite{CANF-VC}    & CANF-based    &  I, P  & N\\
    B-CANF (Ours)             & CANF-based    &  I, B, B* & Y\\
    \hline
    \color{black}
    \end{tabular}
    \label{tab:anf_compare}
\end{table}

\section{Proposed Method}
\label{sec:method}

\subsection{System Overview}
\begin{figure}
    \centering
    \includegraphics[width=0.8\linewidth]{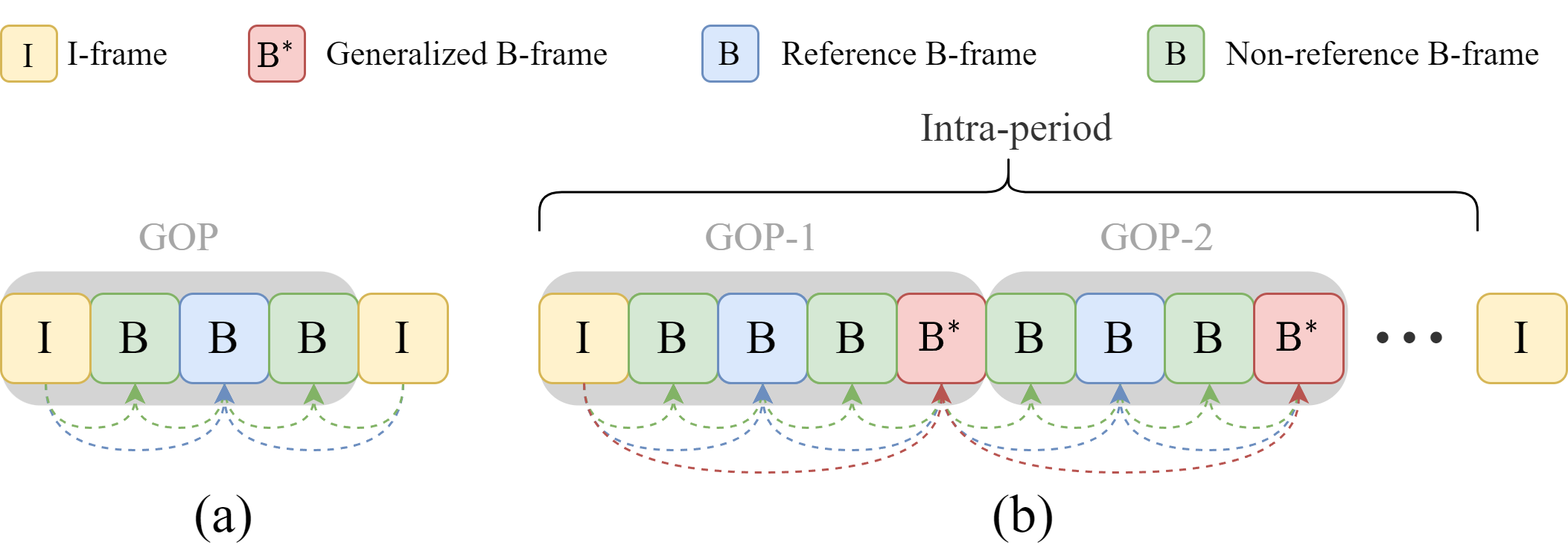}
    \caption{
    \textcolor{black}{
    Illustration of the GOP structures supported in our scheme. The GOP size is set to 4 for illustrative purposes. (a) Hierarchical B-frame coding. (b) Hierarchical B-frame coding with multiple GOPs in an intra-period. In both cases, reference and non-reference B-frames are coded recursively. 
    }}
    \label{fig:coding structure example}
\end{figure}

We begin this section with an overview of our B-frame coding framework. Specifically, we target the two common hierarchical Group-of-Pictures (GOP) prediction structures shown in Fig.~\ref{fig:coding structure example}, where Fig.~\ref{fig:coding structure example}(a) depicts the case with an intra-period of the same size as the GOP and Fig.~\ref{fig:coding structure example}(b) the other case having multiple GOPs in one intra-period. In the latter case, we introduce a special type of B-frame, known as B*-frame, the coding of which reuses the same framework as for regular B-frames (including both reference and non-reference ones). Our B*-frames are conceptually similar to \textit{generalized} B-frames~\cite{h264_generalized_B}, which allow the two reference frames to be both from the past (and even the same) decoded frames. The use of B*-frame is to mimic P-frame coding with our B-frame coding framework, an approach that is in direct contrast to~\cite{BEPIC}, which uses P-frame coding for B-frames. We remark that B-frame coding is more flexible and general than P-frame coding.  

\begin{figure*}[]
    \centering
    \subfloat[]{
        \centering
        \includegraphics[width=0.8\linewidth]{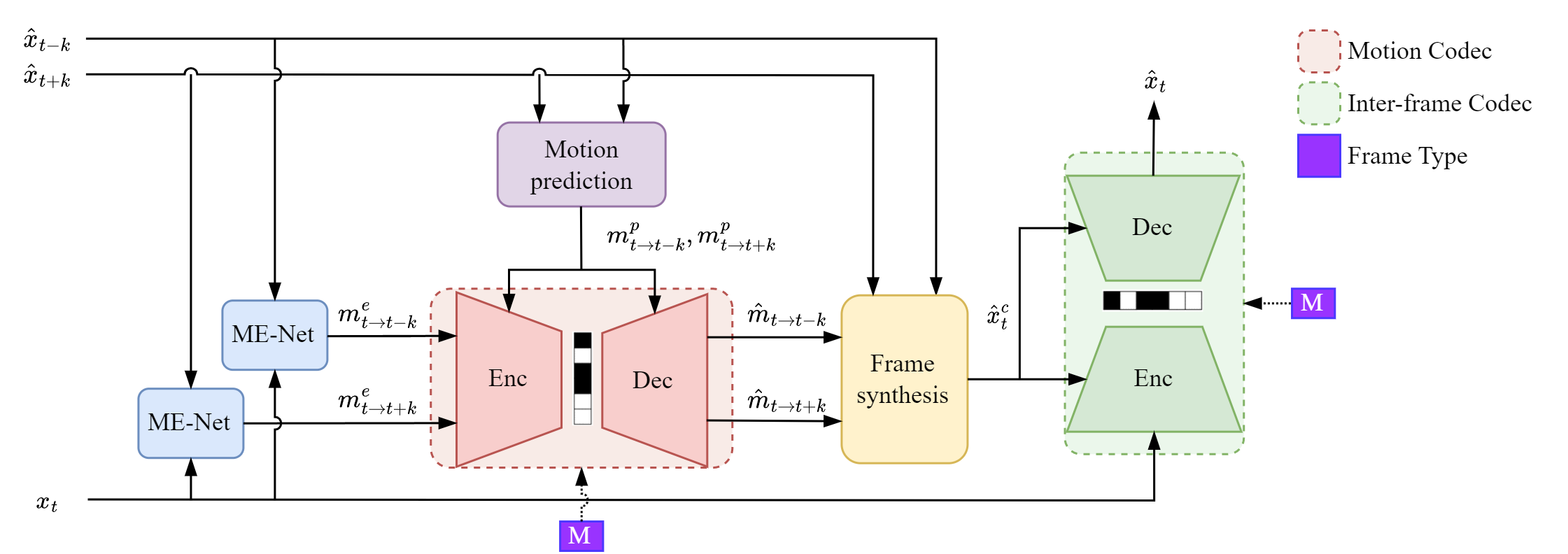}
        \label{fig:overall_a}
    }
    
    \subfloat[]{
        \centering
        \includegraphics[width=0.8\linewidth]{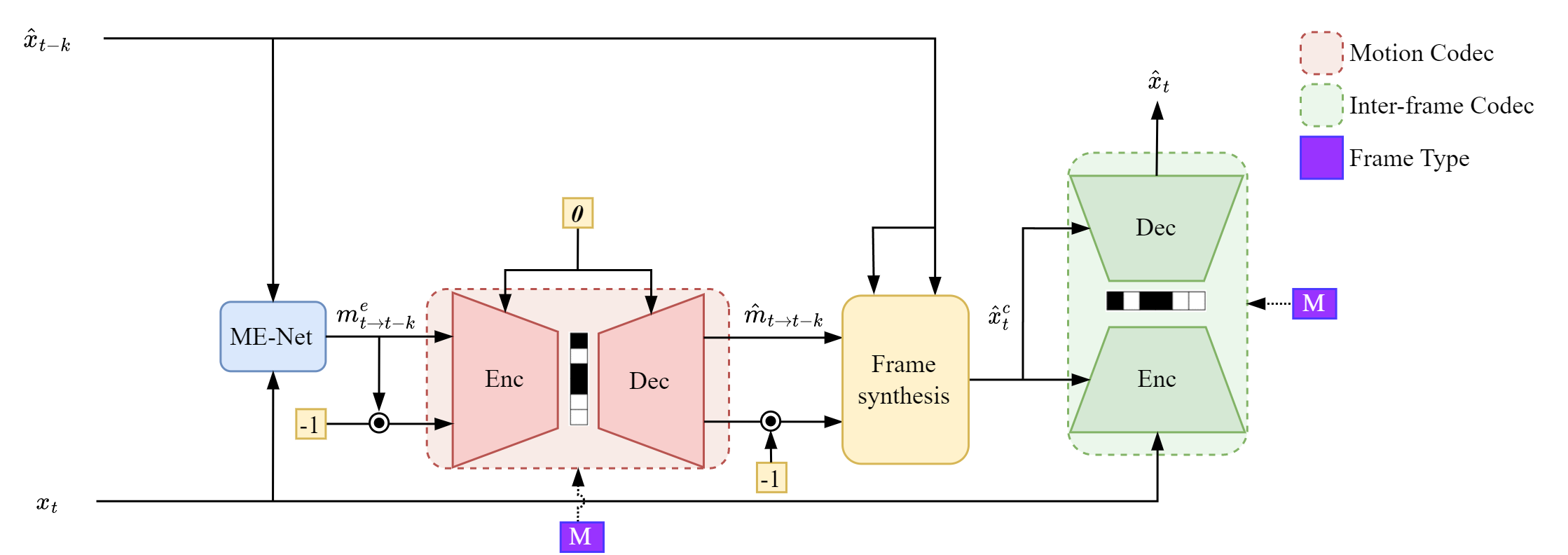}
        \label{fig:overall_b}
    }
    \caption{Illustration of the proposed B-frame codec configured for coding (a) B-frames (reference and non-reference) and (b) \textcolor{black}{B*-frames}. \textcolor{black}{$x_t$ represents the current coding frame and $\hat{x}_{t-k}, \hat{x}_{t+k}$ are the previously reconstructed reference frames. ME-Net estimates the optical flow maps $m^e_{t\to t-k}, m^e_{t\to t+k}$ between $x_t$ and its reference frames $\hat{x}_{t-k}, \hat{x}_{t+k}$, respectively. The motion prediction network outputs the predicted optical flow maps $m^p_{t\to t-k}, m^p_{t\to t+k}$, which serve as the conditioning signals for the motion codec. The frame synthesis network fuses the reference frames using the reconstructed optical flow maps $\hat{m}_{t\to t-k}, \hat{m}_{t\to t+k}$ to generate the predicted frame $\hat{x}^c_t$, which acts as the conditioning signal for the inter-frame codec. $M$ indicates the frame type (reference B, non-reference B, B*-frame).}}
    \label{fig:overall}
\end{figure*}

Fig.~\ref{fig:overall}(a) illustrates our B-frame coding framework. Consider the coding of a B-frame $x_t$, \textcolor{black}{where the subscript indicates the time index}, given its two reference frames $\hat{x}_{t-k},\hat{x}_{t+k}$, i.e., the previous and subsequent decoded reference frames at higher hierarchical levels, respectively. Our first step is to interpolate a predicted frame $\hat{x}_t^c$ through bidirectional motion compensation in pixel domain. To this end, we estimate and signal two optical flow maps $m^{e}_{t \rightarrow t-k},m^{e}_{t \rightarrow t+k}$ for bidirectional backward warping. Our main focus here is on reducing motion overhead through conditional coding. \textcolor{black}{To obtain the conditioning signals, we introduce a motion prediction network that derives the (predicted) optical flow maps $m^{p}_{t \rightarrow t-k},m^{p}_{t \rightarrow t+k}$ from observing the two reference frames $\hat{x}_{t-k},\hat{x}_{t+k}$ without having access to $x_t$.} Our second step is to code the B-frame $x_t$ efficiently by conditioning the inter-frame coding on the predicted frame $\hat{x}_t^c$. Our framework is comprised of the following major components:

\begin{itemize}
   \item \textbf{Motion estimation network (ME-Net)} estimates the backward flow maps $m^{e}_{t \rightarrow t-k},m^{e}_{t \rightarrow t+k}$ between $x_t$ and the two reference frames $\hat{x}_{t-k},\hat{x}_{t+k}$, respectively. In this work, we adopt SPyNet~\cite{spy_net} for motion estimation. 
    
   \item \textbf{Motion prediction network} predicts two optical flow maps $m^{p}_{t \rightarrow t-k},m^{p}_{t \rightarrow t+k}$, which serve jointly as the condition for coding $m^{e}_{t \rightarrow t-k},m^{e}_{t \rightarrow t+k}$, from observing the two reference frames $\hat{x}_{t-k},\hat{x}_{t+k}$ without having access to $x_t$. We follow \cite{huang2020rife} in using a convolutional neural network to generate $m^{p}_{t \rightarrow t-k},m^{p}_{t \rightarrow t+k}$ in a coarse-to-fine fashion. 

   \item \textbf{CANF-based motion codec}, conditioned on the predicted flow maps $m^{p}_{t \rightarrow t-k},m^{p}_{t \rightarrow t+k}$, encodes the flow maps $m^{e}_{t \rightarrow t-k},m^{e}_{t \rightarrow t+k}$ jointly to output the reconstructed flow maps $\hat{m}_{t \rightarrow t-k},\hat{m}_{t \rightarrow t+k}$ (Section~\ref{method:CANF}).
    
   \item \textbf{Frame synthesis network} synthesizes the predicted frame $\hat{x}_t^c$ by GridNet~\cite{gridnet}, which takes as inputs the motion-compensated reference frames and their features generated based on the reconstructed flows $\hat{m}_{t \rightarrow t-k}, \hat{m}_{t \rightarrow t+k}$. 

   \item \textbf{CANF-based inter-frame codec} encodes the target frame $x_t$, conditioned on the predicted frame $\hat{x}_t^c$ (Section~\ref{method:CANF}).

\end{itemize}

\subsection{CANF-based Inter-frame and Motion Codecs}
\label{method:CANF}

Fig.~\ref{fig:anfs}(b) presents the architecture of our CANF-based codec. Both our inter-frame and motion codecs adopt a similar design. To keep the notation uncluttered, we take the inter-frame codec as an illustrative example.

\textbf{CANF-based inter-frame codec} seeks to learn the conditional distribution $p(x_t|\hat{x}_t^c)$ of the B-frame $x_t$, given the bidirectionally predicted frame $\hat{x}_t^c$. As compared to ANFIC~\cite{anfic}, which employs an unconditional ANF to learn the image distribution $p(x_t)$ for image compression, our CANF-based inter-frame codec introduces two novel elements: (1) conditional autoencoding transforms and (2) conditional entropy coding with the combined hyperprior and temporal prior.

From Fig.~\ref{fig:anfs}(b), the conditional autoencoding transform is designed to transform the B-frame $x_t$ into the latent $y_2$, which in the present case is regularized to approximate the predicted frame $\hat{x}_t^c$ instead of a zero image as with ANFIC~\cite{anfic} (cp.~Fig.~\ref{fig:anfs}(a)). In the course of transformation, the latent $\hat{z}_2$ encodes the information necessary for instructing \textcolor{black}{the transformation from $x_t$ to $\hat{x}_t^c$ and vice versa}, while the hyperprior $\hat{h}_2$ is utilized to model the distribution of $\hat{z}_2$ for entropy coding. Taking $\{A_1,S_1\}$ in Fig.~\ref{fig:anfs}(b) as an example, we have 
\begin{align}
    & A_1(x_t,e_z|\hat{x}_t^c)=(x_t, e_z + \mu_{A_1}(x_t, \hat{x}_t^c))=(x_t,z_1), 
    \\
    & S_1(x_t,z_1)=(x_t-\mu_{S_1}(z_1), z_1)=(y_1,z_1), 
\end{align}
where $\mu_{A_1},\mu_{S_1}$ are two neural networks that output element-wise additive transform parameters. It is worth noting that the encoding process $\mu_{A_1}$ is conditioned on $\hat{x}_t^c$ by concatenating $\hat{x}_t^c, x_t$ to form the encoder input. The rationale behind the design is to ease the transformation from $x_t$ to $\hat{x}_t^c$ by supplying the target $\hat{x}_t^c$ as an auxiliary signal. This process is repeated by taking the resulting $y_1,z_1$ as inputs to the next autoencoding transform $\{A_2,S_2\}$.

The conditional entropy coding is achieved by the hierarchical autoencoding transform $\{H_a,H_s\}$, which estimates the conditional distribution of $z_2$ given the hyperprior $\hat{h}_2$ and the interpolated frame $\hat{x}_t^c$ for entropy coding. Its operation is governed by 
\begin{align}
    & H_a(z_2,e_h)=(z_2, e_h + \mu_{H_a}(z_2))=(z_2,\hat{h}_2), 
    \\
    & H_s(z_2,\hat{h}_2|\hat{x}_t^c)=(\lfloor z_2 - \mu_{H_s}(\hat{h}_2, H_p(\hat{x}_t^c))  \rceil, \hat{h}_2)=(\hat{z}_2, \hat{h}_2), %\nonumber
\end{align}
where $\lfloor \cdot \rceil$ (depicted as Q in Fig.~\ref{fig:anfs}(b)) denotes the nearest-integer rounding at inference time for coding $z_2$ in a lossy manner. In particular, $\hat{h}_2=e_h+\mu_{H_a}(z_2)$ denotes the quantized hyperprior and is entropy coded by a learned factorized distribution~\cite{iclr17balle}, with $e_h \sim \mathcal{U}(-0.5, 0.5)$ simulating additive quantization noise applied to the hyperprior latent $\mu_{H_a}(z_2)$. $\mu_{H_s}(\hat{h}_2, H_p(\hat{x}_t^c))$ models the mean of the latent $z_2$, which is assumed to follow a Gaussian distribution with the standard deviation determined by another output $\sigma(\hat{h}_2,H_p(\hat{x}_t^c))$ tied to the same backbone network as $\mu_{H_s}$. We remark that the combination of $\hat{h}_2,\hat{x}_t^c$ in deriving the coding probabilities for $z_2$ exerts a combined effect of the hyperprior and temporal prior. 

In Fig.~\ref{fig:anfs}(b), \textcolor{black}{another conditioning factor $M$ is utilized to signal the frame type of $x_t$--namely, reference B-frames, non-reference B-frames, or B*-frames--and to adapt feature distributions for frame-type adaptive coding (Section~\ref{sec:FrameAdaptCoding}).}

\textbf{CANF-based motion codec} follows a similar design to our CANF-based inter-frame codec. The changes include (1) replacing the coding B-frame $x_t$ with the concatenation of the two optical flow maps $\{m^{e}_{t \rightarrow t-k}, m^{e}_{t \rightarrow t+k}\}$ to be coded, and (2) replacing the interpolated frame $\hat{x}_t^c$ with the concatenation of the two predicted optical flow maps $\{m^{p}_{t \rightarrow t-k}, m^{p}_{t \rightarrow t+k}\}$. It is important to note that $\{m^{e}_{t \rightarrow t-k}, m^{e}_{t \rightarrow t+k}\}$ are coded jointly based on the predicted flows $\{m^{p}_{t \rightarrow t-k}, m^{p}_{t \rightarrow t+k}\}$, allowing the motion codec to exploit correlations between these flow maps for reducing motion overhead.  

\textcolor{black}{The major differences between the CANF-based codecs of our B-CANF and~\cite{CANF-VC} are summarized as follows:}
\begin{itemize}
    \color{black}
    \item \textbf{Bidirectional joint motion coding:} B-CANF addresses B-frame coding. There are two optical flow maps (rather than one in~\cite{CANF-VC}) to be coded. Instead of encoding these flow maps independently, they are concatenated as the input to our CANF-based motion codec for joint coding. Likewise, the motion prediction module outputs two flow map predictors. They are also concatenated as the joint conditioning signal. This allows our CANF-based motion codec to best use the contextual information from both flow map predictors for joint coding. As a result, the channels of the input, conditional input, and output are twice as many as those of their counterparts in~\cite{CANF-VC}.
    \item \textbf{Frame-type adaptive coding:} B-CANF additionally incorporates a frame-type adaption module in every convolutional layer of the CANF-based motion and inter-frame codecs, in order to adapt the feature distributions for frame-type adaptive coding. We will discuss this in Section~\ref{sec:FrameAdaptCoding}.
    \item \textbf{Network configuration:} The network settings, such as the number of input/output channels, kernel size, and stride, for the CANF-based motion and inter-frame codecs have been changed in this work, in order to strike a better balance among coding efficiency, model size, and computational complexity. More details will be provided in the released code.
    \color{black}        
\end{itemize}

\subsection{B*-frame Extension} \label{sec:B_frame_extension} 
\textcolor{black}{B*-frames allow to support multiple GOPs in an intra-period as depicted in Fig.~\ref{fig:coding structure example}(b), to strike a balance between coding performance and latency. A straightforward approach is to encode P-frames periodically using an additional P-frame codec. In doing so, however, the total model size is increased. To address this issue, we propose B*-frames, which reuse our CANF-based B-frame codec to mimic P-frame coding.} 

To encode a video frame $x_t$ in B* mode, we create a predicted frame from the past decoded frame $x_{t-k}$ just like coding a vanilla P-frame. To this end, the flow map $m^e_{t\to t-k}$ between $x_t$ and $x_{t-k}$ is estimated and coded. As shown in Fig.~\ref{fig:overall}(b), the main inputs to our conditional motion codec should ideally comprise two optical flow maps. We generate the other "virtual" input by flipping the sign of $m^e_{t\to t-k}$ (i.e. $-1 \times m^e_{t\to t-k})$, in order to match the characteristics of $m^e_{t\to t+k}$ (for ordinary B-frame coding). \textcolor{black}{Because only one reference frame $x_{t-k}$ is used, we disable motion prediction network and set the predicted flow maps to constant 0. This reduces our motion coding scheme for B*-frames to intra-based motion coding. Although it is possible to use a motion extrapolation network to generate a flow map predictor from previously coded B*-frames, this calls for long-term motion extrapolation. It is to be noted that B*-frames are usually temporally distant from each other or may even be coded only once in an intra-period as with the HM randomaccess configuration~\cite{HM}.} The predicted frame $\hat{x}_t^c$ is then synthesized 
% by feeding into the frame synthesis network the warped reference frames
\textcolor{black}{from the reference frames}
based on the coded $m^e_{t\to t-k}$ and its virtual counterpart (after sign reversal), which are found empirically to be similar but not exactly the same (see ablation experiments in Section~\ref{sec:ablation}). As such, the B*-frame is by definition a bi-prediction frame. 

\textcolor{black}{Our ablation experiments in Section~\ref{sec:ablation} show that B*-frames have similar compression performance to P-frames, which require an additional, separate P-frame codec.}

\subsection{Frame-type Adaptive Coding}
\label{sec:FrameAdaptCoding}
Frame-type adaptive coding aims to achieve adaptive coding according to the reference types of B-frames. In traditional codecs, the \textit{reference} B-frames are usually coded at higher quality than the \textit{non-reference} B-frames by operating the same B-frame codec in different modes\footnote{The reference B-frames refer to those B-frames which serve as reference frames for the subsequent frames in coding order, whereas the non-reference B-frames 
are not used for reference.}. \textcolor{black}{Following a similar strategy, we weight more heavily the distortions of the reference B-frames and B*-frames during training (Section~\ref{sec:3.4}). In addition, we observe that using a fixed, shared B-frame codec for coding all the types of B-frames is unable to achieve frame-type adaptive coding.} On the other hand, training separate models for different B-frames as with~\cite{Wu_2018} is prohibitively expensive. 

\begin{figure}
    \centering
    \includegraphics[width=\linewidth]{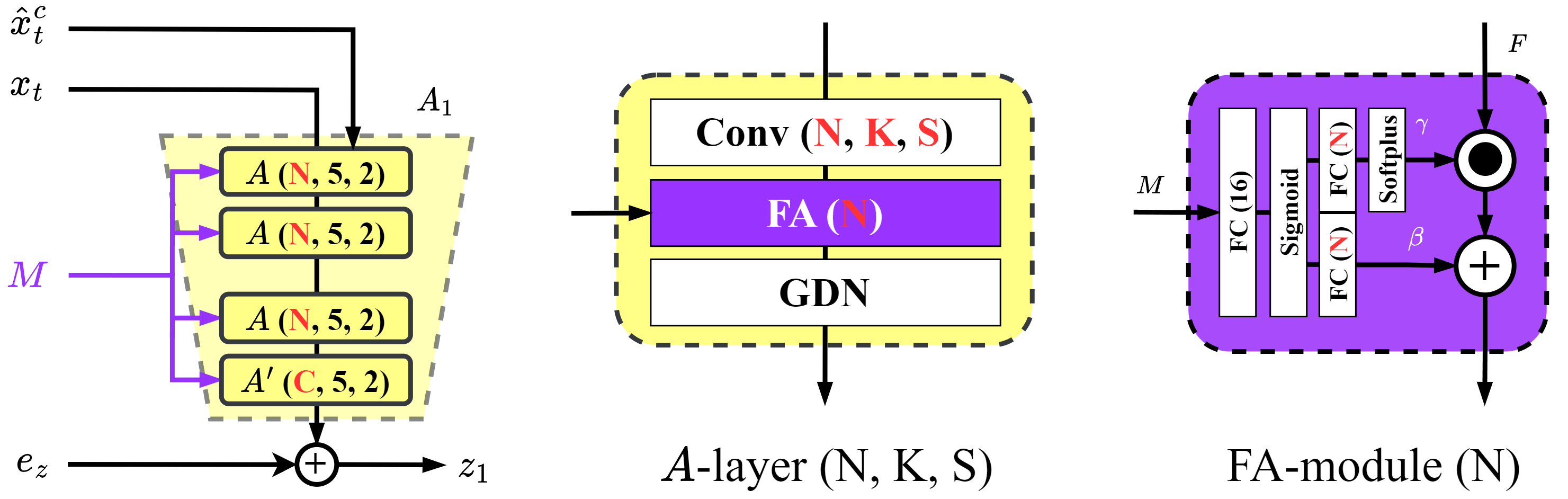}
    \caption{\textcolor{black}{Illustration of the analysis transform ($A_1$) with the frame-type adaptation (FA) module. The symbol $'$ indicates that the activation function (i.e. GDN) is omitted. Conv(N, K, S) stands for a convolutional layer with the number of output channels N, kernel size K, and stride S.}}
    \label{fig:FA_module}
\end{figure}

We tackle this issue by introducing a frame-type adaptation (FA) module to every convolutional layer, aiming to share the convolutional layers between B-frames while adapting their features to the B-frame type. Specifically, according to the B-frame type (namely, reference B-frames, non-reference B-frames, or B*-frames), our FA module applies a channel-wise affine transformation to the output features of every convolutional layer. Take as an example the analysis transform in Fig.~\ref{fig:FA_module}. The FA module is inserted between the convolutional layer and the generalized divisive normalization (GDN) layer. It takes B-frame type $M$ (in one-hot vector) as input and generates the affine transform parameters $\gamma$ and $\beta$ for feature $F$ adaptation by
\begin{align}
    \text{FA}(F|M) =\gamma(M)\odot F\oplus\beta(M)
\end{align}
where $\odot$, $\oplus$ are channel-wise multiplication and addition. The FA modules appear in the autoencoding transforms and the hyperprior models. They are trained end-to-end together with the other networks. \textcolor{black}{Note that similar feature adaptation techniques are widely used in computer vision tasks~\cite{GateAdapt, SFT} and are also adopted in end-to-end learned image compression~\cite{SFTImageCoding}. Unlike these prior works, which take a set of feature maps as inputs to perform element-wise affine transformation, our FA module performs channel-wise affine transformation to adapt feature distributions with respect to a scalar indicating the input frame's type.}

Section~\ref{sec:ablation} shows that frame-type adaptive coding improves bit allocation because our B-frame codec is able to adapt feature distributions on a frame-by-frame basis. In passing, this tool feature is to be distinguished from training separate models for different rate points, which is still the case in all the experiments.

\subsection{Training}
\label{sec:3.4}
The model training is done on 5-frame training sequences (with their frames denoted by $x_0,x_1, \ldots ,x_4$). In coding every training sequence, we follow the GOP-1 structure in Fig.~\ref{fig:coding structure example}(b), which encodes $x_0$ as an I-frame by a pre-trained image codec~\cite{anfic} (not included for training), $x_4$ as a B*-frame\textcolor{black}{, and $x_2$ as a reference B-frame.} Moreover, we randomly choose between $x_1$ and $x_3$ for coding, in order to have a balanced distribution over B-frame types. Assuming that $x_3$ is chosen, we formulate our training objective as follows:
\begin{align}
    \mathcal{L} &= \lambda_1 \times D + R + \lambda_2 \times F,
    \label{eq:rdloss}
\end{align} 
\textcolor{black}{where
\begin{align}
   D &= d(x_4, \hat{x}_4) + \frac{1}{\alpha_{r}} d(x_2, \hat{x}_2) + \frac{1}{\alpha_{nr}} d(x_3, \hat{x}_3), \\
   R &= \sum_{t=2}^{4}  r(\hat{x}_t),  \\
   F &= f_4 + \frac{1}{\alpha_{r}} f_2 + \frac{1}{\alpha_{nr}} f_3,
   % \mathcolor{black}{f_t &= ||y_2 - \hat{x}^c_t||^2_2},
\end{align} 
}
\textcolor{black}{with $D$ denoting the weighted sum of frame distortions $\{d(x_t,\hat{x}_t)\}_{t=2}^4$, $R$ the accumulated bits $\{r(\hat{x}_t)\}_{t=2}^4$ consumed by both the conditional motion and inter-frame coding, and $F$ the sum of regularization losses \textcolor{black}{$\{f_t=||y_2 - \hat{x}^c_t||^2_2\}_{t=2}^4$} requiring that $y_2$ in Fig.~\ref{fig:anfs}~(b) should approximate their respective conditions, i.e. $\{\hat{x}_t^c\}_{t=2}^4$. \textcolor{black}{This regularization loss is to have the latent representations $\hat{z}_2,\hat{h}_2$ (Fig.~\ref{fig:anfs}~(b)) capture the information needed to signal the transformation between the coding frame $x_t$ and its predicted frame $\hat{x}^c_t$}. The hyper-parameters $\lambda_1,\lambda_2$ are rate-point dependent, while $\alpha_{r}, \alpha_{nr}$ are for bit allocation among B-frames. These hyper-parameters are detailed in Section~\ref{sec:experiment}.
}
\section{Experiments}
\label{sec:experiment}

\subsection{Settings}
    \paragraph{Training Details}
    \label{sec:training}
    For training, we use Vimeo-90k septuplet dataset~\cite{vimeo}, which includes 91,701 7-frame sequences of size $448 \times 256$. The training sequences are randomly cropped to $256 \times 256$, and five consecutive frames are selected. We adopt Adam~\cite{adam} optimizer with learning rate $10^{-4}$ and batch size 16. \textcolor{black}{To optimize our models for PSNR-RGB, we choose \textcolor{black}{$\lambda_2=0.01\lambda_1 \text{ and } \lambda_1={128,512,1024,2048}$} to train separate models for different rate points; likewise, to optimize them for MS-SSIM-RGB, we choose \textcolor{black}{$\lambda_2=0.01\lambda_1 \text{ and } \lambda_1={2,8,16,32}$}. For all the experiments, $\alpha_{r}=1, \alpha_{nr}=2$. These hyper-parameters are chosen to weight more heavily the quality of reference B-frames (including B*-frames).} 
    
\begin{table*}[t]
    \centering
    \setlength\tabcolsep{2.9pt}
    \caption {Settings of the competing methods.}
    \vspace{\baselineskip}
    \begin{tabular}{lccccc}
        \Xhline{2\arrayrulewidth}
                & I-frame codec & Coding & Intra- & \multirow{2}{*}{GOP}\\
                & (PSNR-RGB/MS-SSIM-RGB) &  structure& period &          \\
        \Xhline{2\arrayrulewidth}
        x265 (veryslow)          & -                           &   IPPP...     & 32 & 32  \\
        HM (LDP)                 & -                           &   Hierarchical P     & 32 & 8       \\
        HM (randomaccess)        & -                           &   Hierarchical B     & 32 & 16       \\
        VTM (randomaccess)       & -                           &   Hierarchical B     & 32 & 32       \\\hline
        DCVC~\cite{dcvc}         & cheng2020~\cite{compressai}/hyperprior~\cite{googleiclr18} & IPPP... & 32 & 32      \\
        DCVC (ANFIC)             & ANFIC~\cite{anfic}/ANFIC              & IPPP... & 32 & 32      \\
        \textcolor{black}{CANF-VC~\cite{CANF-VC}}   & ANFIC/ANFIC        & IPPP... & 32 & 32      \\
        Sheng'22~\cite{tcm}      & cheng2020~\cite{compressai}/cheng2020 & IPPP... & 32 & 32 \\
        Sheng'22 (ANFIC)     & ANFIC~\cite{anfic}/ANFIC & IPPP... &     32 & 32 \\
        Li'22~\cite{Li_2022}     & Li'22~\cite{Li_2022}/Li'22            & IPPP... & 32 & 32 \\
        Li'22 (ANFIC)   & ANFIC~\cite{anfic}/ANFIC & IPPP... & 32 & 32 \\
        LHBDC~\cite{murat_lhbdc} & mbt2018-mean~\cite{compressai}/-   & Hierarchical B & 32 & 16      \\
        LHBDC (ANFIC)            & ANFIC/-                & Hierarchical B & 32 & 16      \\\hline
        \textbf{B-CANF (Ours)}          & ANFIC/ANFIC                 & Hierarchical B & 32 & 16      \\
        \Xhline{2\arrayrulewidth}
    \end{tabular}
    \label{tab:setting}
\end{table*}

\begin{table*}[t]
    \centering
    \setlength\tabcolsep{2.0pt}
    \caption {BD-rate comparison with intra-period 32. The anchor is x265 in veryslow mode. The best performer is marked in \textcolor{red}{red} and the second best in \textcolor{blue}{blue}.}
    \vspace{\baselineskip}
        
    \begin{tabular}{lccccc@{}p{0.5em}@{}cccc}
        \Xhline{2\arrayrulewidth}
        & \multicolumn{4}{c}{BD-rate (\%) PSNR-RGB} && & \multicolumn{4}{c}{BD-rate (\%) MS-SSIM-RGB}   \\
        \cline{2-5} \cline{8-11}
                                 &  UVG  & MCL-JCV & HEVC-B & CLIC'22 &&& UVG  & MCL-JCV & HEVC-B & CLIC'22 \\
        \Xhline{2\arrayrulewidth}
        
        HM (LDP)                 &  -30.6&  -26.4  &  -23.0 &  -16.8  &&& -21.6&  -17.3  &  -18.0 & -8.8    \\ 
        HM (randomaccess)        &  -53.6&  -48.2  &  -51.4 & -41.3   &&& -46.3&  -40.8  &  -53.3 & -39.2   \\
        VTM (randomaccess)   &  \color{red}-68.2 &  \color{red}-63.6  &  \color{red}-66.4 &  \color{red}-54.6  &&& -69.6 &  -65.6  &  -71.1 & -60.9 \\\hline
        DCVC~\cite{dcvc}         &  -2.3 &  -7.6   &   -3.0 &  28.7   &&& -45.2&  -49.5  &  -53.6 & -25.9   \\
        DCVC (ANFIC)             &  -3.6 &  -9.0   &   -2.3 &  25.5   &&& -47.7&  -51.6  &  -55.8 & -29.0   \\
        CANF-VC~\cite{CANF-VC}   &  -28.7&  -21.3  &   -21.4&  3.3    &&& -49.8&  -50.4  &  -55.0 & -30.3   \\
        Sheng'22~\cite{tcm}      & -45.7 & -38.0    & -43.1  & -9.5   &&& -63.6& -63.8 & -73.9 & -58.2 \\
        Sheng'22 (ANFIC)         & -41.8 & -38.0    & -40.9 & -9.9   &&& -63.0& -64.7 & -73.3 & -57.1 \\
        Li'22~\cite{Li_2022}     & \textcolor{blue}{-62.4} & \textcolor{blue}{-54.0}    & \textcolor{blue}{-57.6}  & -26.9  &&& \textcolor{red}{-73.0} & \textcolor{red}{-72.8} & \textcolor{red}{-80.4} & \textcolor{red}{-67.9} \\
        Li'22 (ANFIC)            & -59.2 & -51.2    & -54.7  & -26.3  &&& \textcolor{blue}{-72.1} & \textcolor{blue}{-71.9} & \textcolor{blue}{-79.6} & \textcolor{blue}{-66.6} \\
        LHBDC~\cite{murat_lhbdc} &  -18.6&  -5.3   &    4.4 &  38.8   &&& NA  &  NA    & NA    & NA     \\
        LHBDC (ANFIC)            &  -18.6&  -6.7   &    3.4 &  38.2   &&& NA  &  NA    & NA    & NA     \\\hline
        \textbf{B-CANF (Ours)}   &  -47.5&  -41.1  &  -46.9 & \textcolor{blue}{-33.8}   &&& -60.2 & -61.2 & -67.1 & -59.6 \\
        \Xhline{2\arrayrulewidth}
    \end{tabular}
    \label{tab:BDrate_gop32}
\end{table*}

    \paragraph{Evaluation Methodologies}
    We evaluate our method on UVG~\cite{uvg}, MCL-JCV~\cite{mcl}, and HEVC Class B~\cite{hevcctc}. In addition, we also evaluate on CLIC'22 test dataset~\cite{CLIC22}, which contains some challenging sequences with characteristics rarely seen in Vimeo-90k training dataset~\cite{vimeo}. To compare fairly with HM (HEVC Test Model Version 16.23)~\cite{HM} under the common test conditions, we follow the coding structure of the \textit{encoder\_randomaccess\_main} configuration. That is, the intra-period is 32 and GOP size is 16 for all the test sequences. For every test sequence, we encode \textcolor{black}{most of the} video frames to the extent that the coded sequence contains the maximum number of intra-periods (32). As an example, for a 600-frame test sequence, we will encode 577 frames (1 I-frame + 18 intra-periods). \textcolor{black}{Note that the source videos are in YUV420 format. The traditional codecs operate internally in YUV420, whereas the learned codecs perform YUV420-to-RGB444 conversion prior to coding.} The reconstructed quality is measured in PSNR-RGB and MS-SSIM-RGB, and the bit-rate in bits-per-pixel (bpp). Table~\ref{tab:setting} summarizes the settings of these competing methods.

    \paragraph{Baseline Methods}
    \label{exp:baseline}
    
    The baseline methods include traditional codecs and learning-based methods. The traditional codecs are x265~\cite{x265} in \textit{veryslow} mode (zerolatency), HM~\cite{HM} with the \textit{encoder\_lowdelay\_P\_main} and \textit{encoder\_randomaccess\_main} configurations, and VTM~\cite{vtm} with the \textit{encoder\_randomaccess\_main} configuration.
    The learning-based methods include CANF-VC~\cite{CANF-VC}, DCVC~\cite{dcvc} and two concurrent works, Sheng'22~\cite{tcm} and Li'22~\cite{Li_2022}, which are the state-of-the-art learned P-frame coding schemes. We also compare B-CANF with LHBDC~\cite{murat_lhbdc}, the state-of-the-art learned B-frame coding scheme.
    To align the I-frame codec, we additionally evaluate their performance using ANFIC~\cite{anfic} as the I-frame codec. We refer to these variants as DCVC (ANFIC), Sheng'22 (ANFIC), Li'22 (ANFIC) and LHBDC (ANFIC), respectively.
    Results for the baselines and their variants are produced with the code released by the authors.

\subsection{Rate-Distortion Performance}

When comparing our B-CANF with CANF-VC~\cite{CANF-VC} and LHBDC~\cite{murat_lhbdc}, we observe that B-CANF surpasses them in terms of both PSNR-RGB and MS-SSIM-RGB across all the test datasets. However, B-CANF falls short of achieving the same level of performance as Li'22~\cite{Li_2022}, which is considered the state-of-the-art learned P-frame codec. This discrepancy in performance can be attributed to the more advanced entropy coding model employed by Li'22~\cite{Li_2022} and the domain shift issue that arises between the training and testing phases. For a comprehensive analysis of this domain shift, please refer to Section~\ref{sec:domain_shift}.

\begin{figure*}[t!]
        \hspace{-1.2em}
        % \subfloat[UVG, PSNR-RGB]{
        \subfloat[]{
            \centering
            \includegraphics[width=0.24\linewidth]{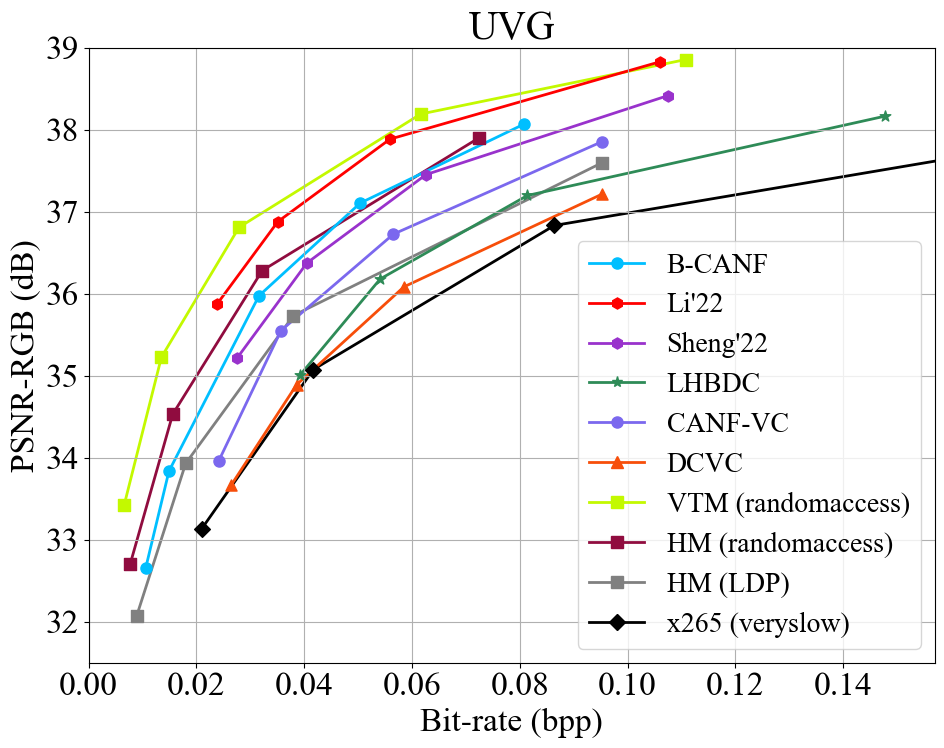}
        }
        % \subfloat[MCL-JCV, PSNR-RGB]{
        \subfloat[]{
            \centering
            \includegraphics[width=0.24\linewidth]{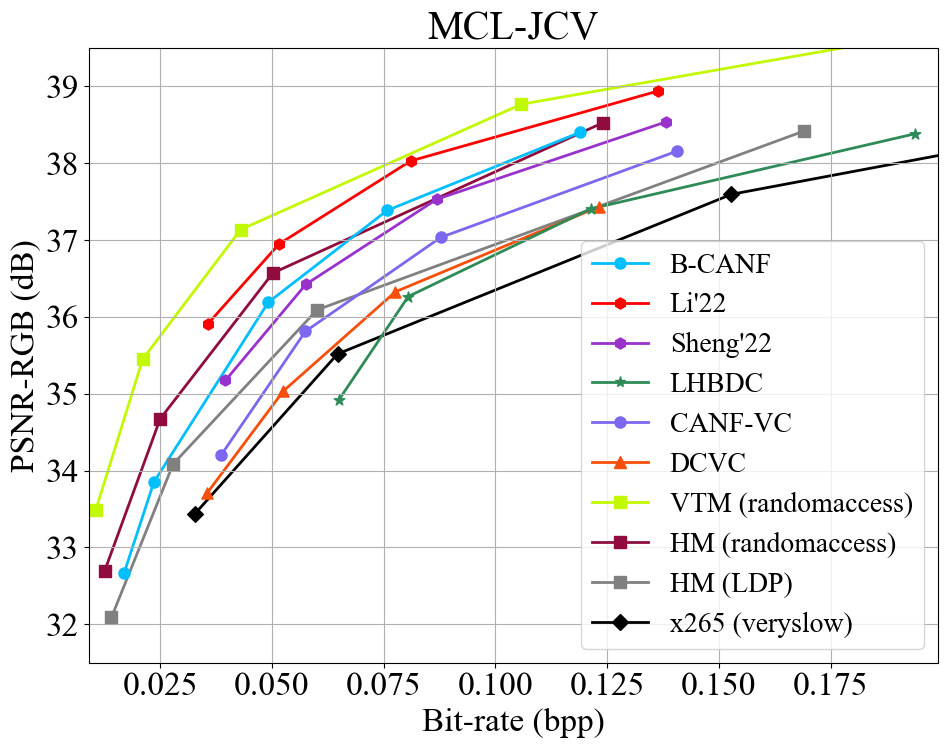}
        }
        % \subfloat[HEVC-B, PSNR-RGB]{
        \subfloat[]{
            \centering
            \includegraphics[width=0.24\linewidth]{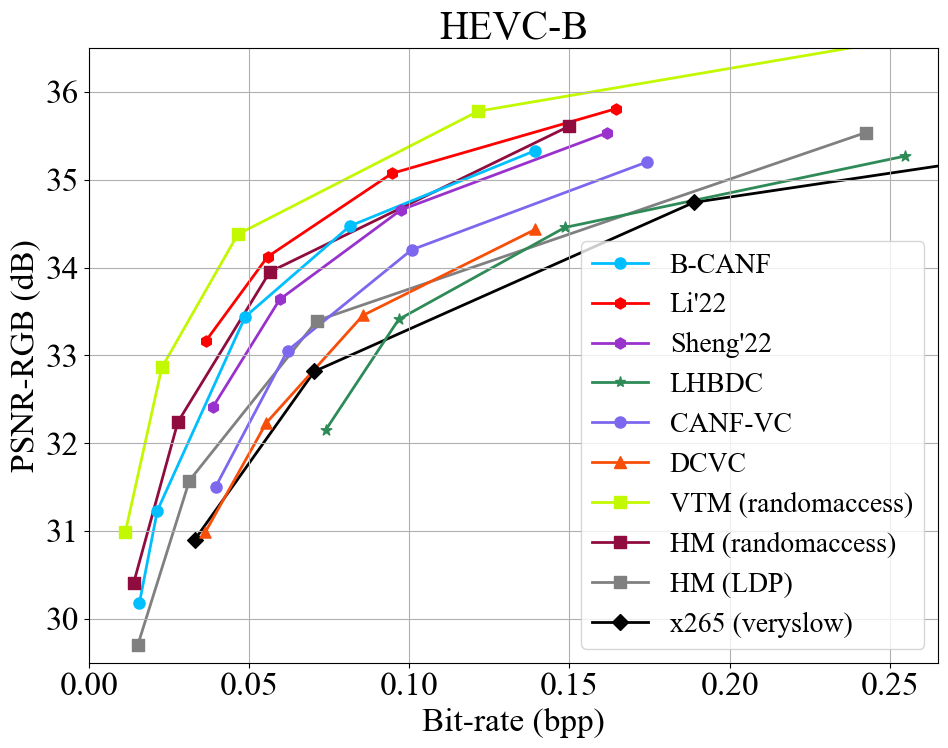}
        }
        % \subfloat[CLIC'22, PSNR-RGB]{
        \subfloat[]{
            \centering
            \includegraphics[width=0.24\linewidth]{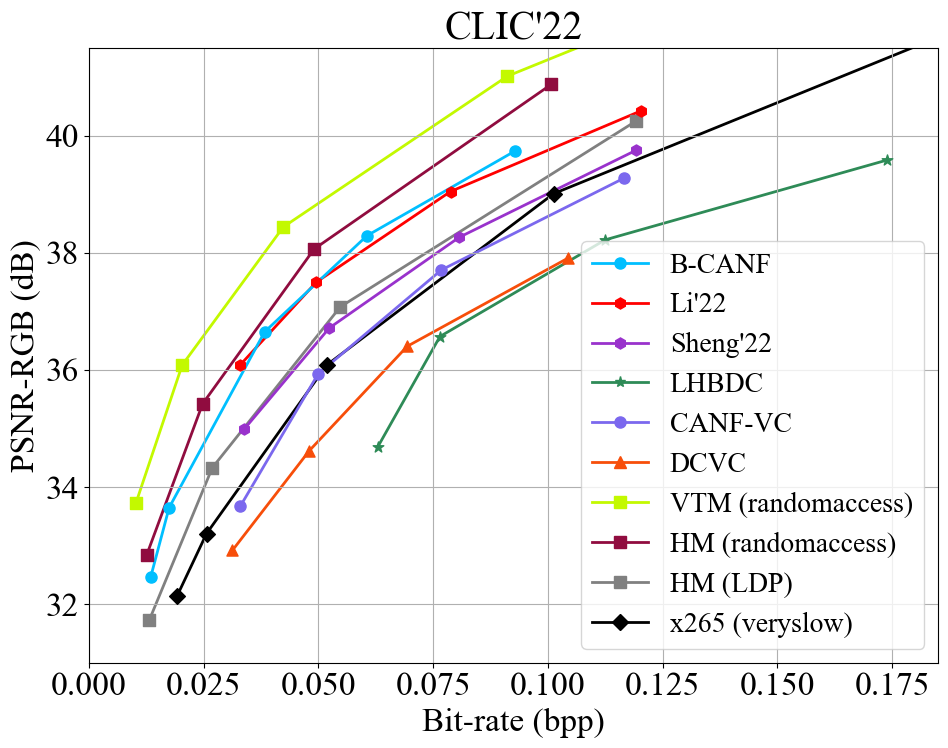}
        }
        
        \hspace{-1.2em}
        % \subfloat[UVG, MS-SSIM-RGB]{
        \subfloat[]{
            \centering
            \includegraphics[width=0.24\linewidth]{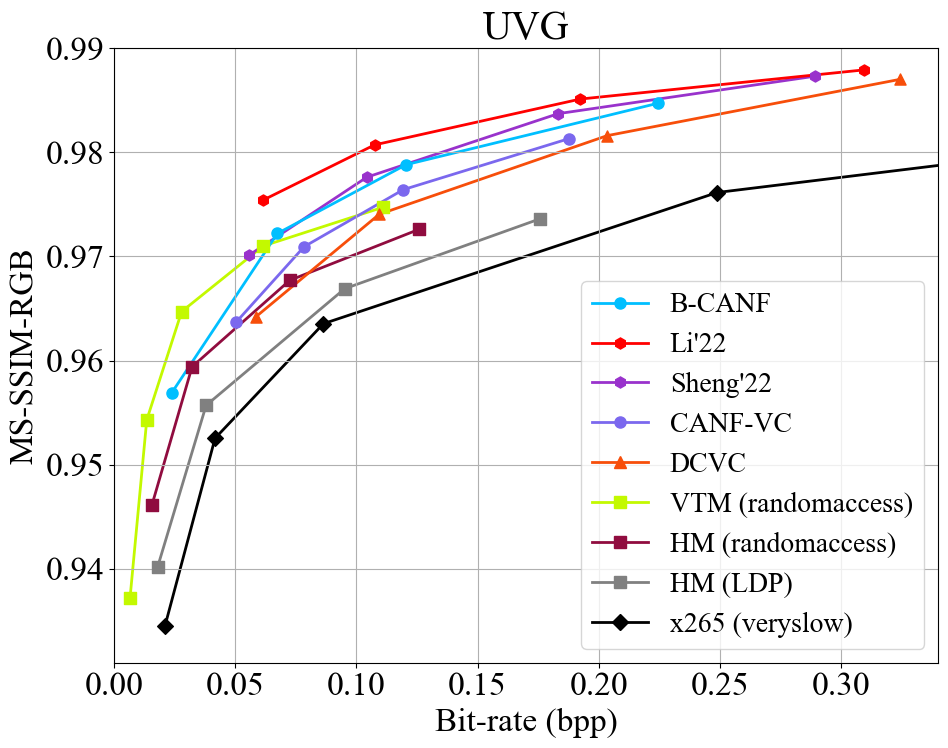}
        }
        % \subfloat[MCL-JCV, MS-SSIM-RGB]{
        \subfloat[]{
            \centering
            \includegraphics[width=0.24\linewidth]{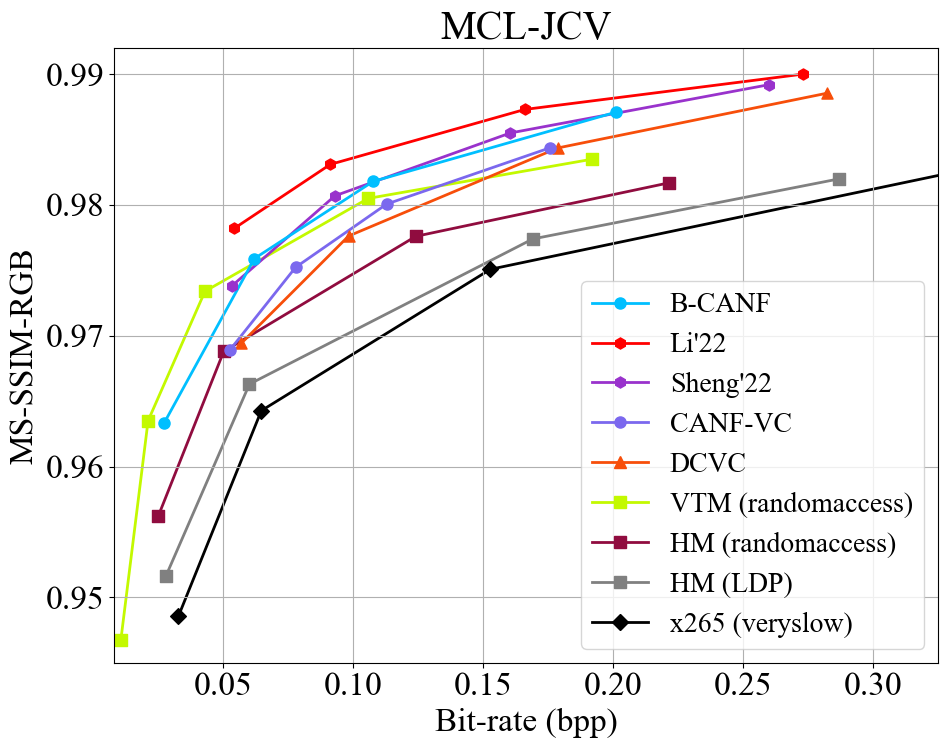}
        }
        % \subfloat[HEVC-B, MS-SSIM-RGB]{
        \subfloat[]{
            \centering
            \includegraphics[width=0.24\linewidth]{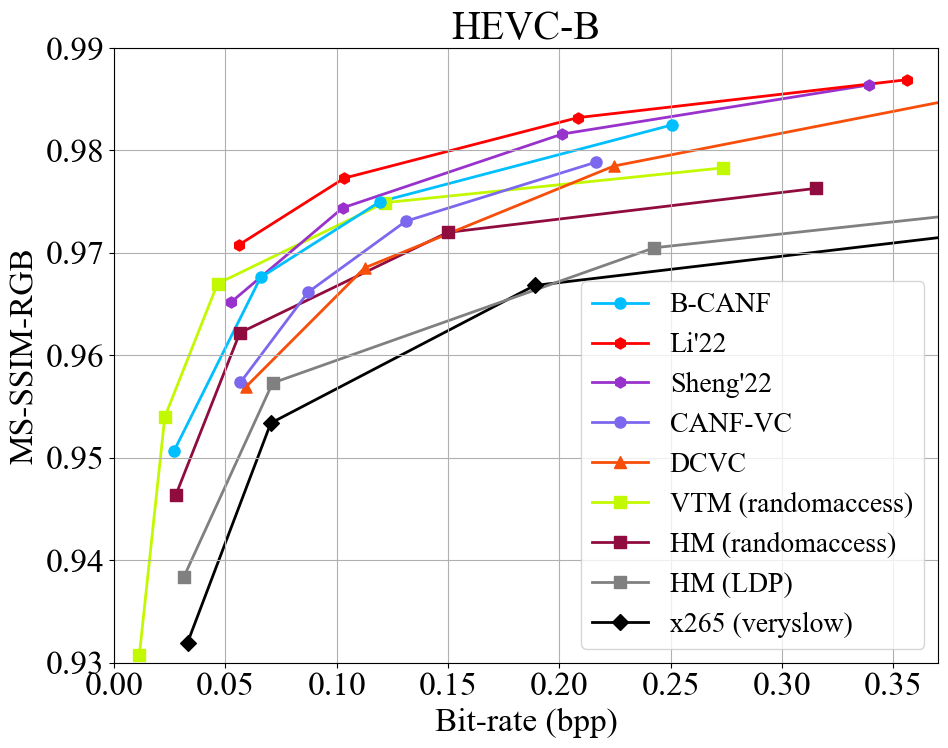}
        }
        % \subfloat[CLIC'22, MS-SSIM-RGB]{
        \subfloat[]{
            \centering
            \includegraphics[width=0.24\linewidth]{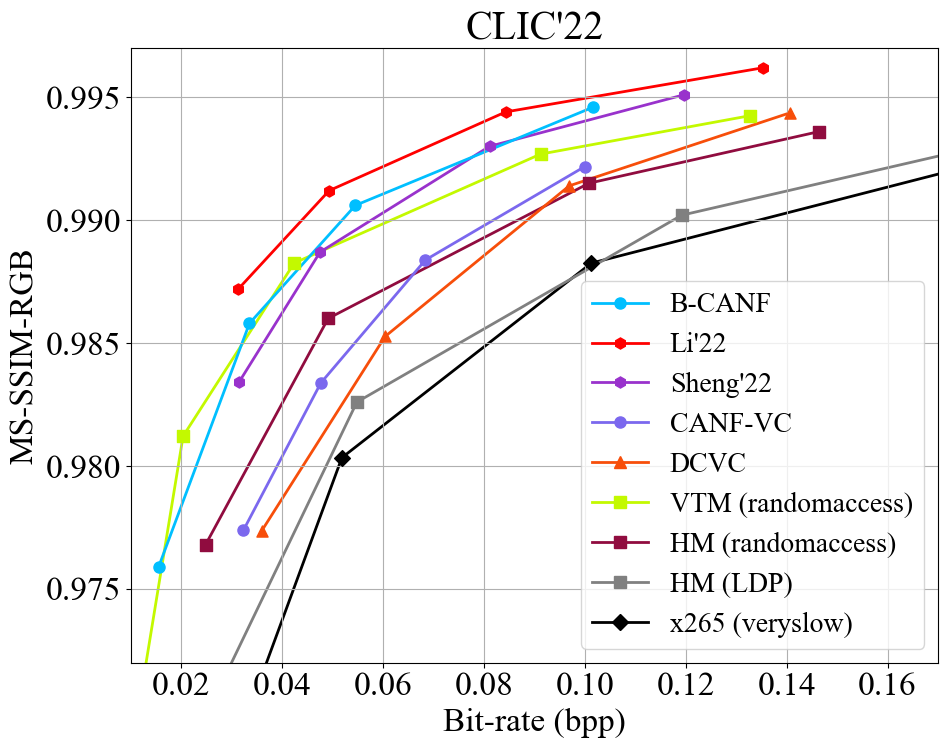}
        }
    \caption{Rate-distortion plots on UVG, MCL-JCV, HEVC Class B and CLIC'22 test dataset in terms of PSNR-RGB and MS-SSIM-RGB}
    \label{fig:psnr_bitrate}
\end{figure*}

\begin{table*}[t]
    \centering
    \setlength\tabcolsep{2.9pt}
    \caption {BD-rates of B-CANF as compared against different anchors under their respective settings.} 
        
    \begin{tabular}{lccc@{}p{0.5em}@{}cccc@{}p{0.5em}@{}cccc}
        \Xhline{2\arrayrulewidth}
        \multirow{2}{*}{Anchor} & I-frame codec & \multirow{2}{*}{Intra-period} & \multirow{2}{*}{GOP}&& \multicolumn{4}{c}{BD-rate (\%) PSNR-RGB}&& \multicolumn{4}{c}{BD-rate (\%) MS-SSIM-RGB} \\
        \cline{6-9}\cline{11-14} &(PSNR-RGB/MS-SSIM-RGB)&&&&  UVG  & MCL-JCV & HEVC-B & CLIC'22 &&  UVG  & MCL-JCV & HEVC-B & CLIC'22\\
        \Xhline{2\arrayrulewidth}
        
        HLVC~\cite{hlvc}         & BPG~\cite{bpg}/ICLR'19~\cite{Lee2019Context}                  & 10        & 10 && -53.7 & -55.4 & -50.3 & -56.9 && -44.2 & -48.5 & -38.7 & -44.8 \\
        HLVC (ANFIC)             & ANFIC~\cite{anfic}/ANFIC                                      & 10        & 10 && -53.9 & -54.9 & -50.3 & -56.9 && -44.7 & -49.0 & -39.0 & -44.1 \\
        RLVC~\cite{rlvc}         & BPG/ICLR'19                                                   & 13        & 13 && -50.6 & -49.4 & -46.5 & -47.8 && -36.6 & -41.9 & -33.6 & -48.9 \\
        RLVC (ANFIC)             & ANFIC/ANFIC                                                   & 13        & 13 && -47.8 & -46.9 & -44.1 & -46.9 && -34.0 & -39.3 & -35.3 & -46.1 \\
        LHBDC~\cite{murat_lhbdc} & mbt2018-mean~\cite{compressai}/mbt2018-mean & 8         & 8  && -33.6 & -36.1 & -43.1 & -41.2 && -31.4 & -     & -     & - \\ 
        LHBDC (ANFIC)            & ANFIC/ANFIC                    & 8         & 8  && -30.9 & -32.9 & -40.7 & -39.9 && -     & -     & -     & - \\  
        B-EPIC~\cite{BEPIC}      & NA/NA                                & infinity  & 12 && -32.3 & -33.8 & -32.5 & -     && -15.5 & -21.2 & -26.5 & - \\
        \Xhline{2\arrayrulewidth}        
    \end{tabular}
    \label{tab:compare}
    \vspace{-1em}
\end{table*}
Table~\ref{tab:compare} further compares our method with \textcolor{black}{HLVC~\cite{hlvc}, RLVC~\cite{rlvc},} LHBDC~\cite{murat_lhbdc}, and B-EPIC~\cite{BEPIC} following their suggested settings. All the baseline methods adopt or support B-frame coding. In this experiment, these baseline methods are used as anchors for BD-rate evaluation. In addition, we note that the codes for B-EPIC~\cite{BEPIC} are unavailable. We thus copy the results from their paper in evaluating BD-rates.

\color{black} 
In HLVC~\cite{hlvc}, a three-level hierarchical coding structure with P- and B-frames is used for encoding GOP's of size 10. We follow exactly the same coding structure as HLVC~\cite{hlvc} to perform encoding with B-CANF, except that we use B*-frames to replace P-frames. Recall that B*-frames are used to mimic P-frames. In RLVC~\cite{rlvc}, a bidirectional P-frame coding structure is proposed to take advantage of the past and future frames for better coding efficiency. For the same purpose, our B-CANF proposes B-frame coding. To demonstrate the best achievable performance of B-CANF, we adopt hierarchical B-frame coding for encoding GOP's of size 13 in comparing with RLVC~\cite{rlvc}. 
In LHBDC~\cite{murat_lhbdc}, a hierarchical B prediction structure with GOP 8 and intra-period 8 is proposed. In comparison with B-EPIC~\cite{BEPIC}, which adopts a GOP size of 12 with an infinite intra-period (only one I-frame in the beginning), our B-CANF uses the same GOP size. However, we set the intra-period to 12 (instead of infinity) to mitigate the temporal error propagation. Despite that this leads to more frequent coding of I-frames, its advantages outweigh the temporal error propagation. Under these various settings, our B-CANF (which adopts ANFIC~\cite{anfic} as the I-frame codec) still achieves considerable BD-rate reductions over HLVC~\cite{hlvc}, RLVC~\cite{rlvc}, LHBDC~\cite{murat_lhbdc}, and B-EPIC~\cite{BEPIC}. This is evidenced by the negative BD-rate numbers in the upper half of Table~\ref{tab:compare}.

\subsection{Domain Shift in Training B-frame Codecs}
\label{sec:domain_shift}

The performance gap between our B-CANF and the works of Sheng'22~\cite{tcm} and Li'22~\cite{Li_2022} deserves further investigation. A crucial aspect that is often overlooked when comparing the compression performance of learned P-frame and B-frame codecs is the training datasets. Most learned codecs (P-frame and B-frame) are trained on Vimeo dataset~\cite{vimeo}, which consists of short video sequences, each containing only 7 frames. Consequently, we use 5-frame GOPs with 3-level B-frame coding for training. Notably, these training B-frames are temporally adjacent to each other.
\begin{table}[t]
    \centering
    \setlength\tabcolsep{2.0pt}
    \caption {BD-rate comparison under different intra-periods: intra-period 32 (I32) and intra-period 8 (I8). The anchor is our proposed method, B-CANF, under the respective intra-periods. Negative numbers indicate rate savings as compared to B-CANF, while positive numbers indicate rate inflation.}
    \vspace{\baselineskip}
        
    \begin{tabular}{lcc@{}p{0.5em}cc@{}p{0.5em}cc}
        \Xhline{2\arrayrulewidth}
        & \multicolumn{8}{c}{BD-rate (\%) PSNR-RGB} \\
        \cline{2-9}
                                 &  \multicolumn{2}{c}{UVG}  && \multicolumn{2}{c}{MCL-JCV} && \multicolumn{2}{c}{HEVC-B}\\
        \cline{2-3} \cline{5-6} \cline{8-9}                       
                                 &  I32 & I8 && I32 & I8 && I32 & I8\\
        \Xhline{2\arrayrulewidth}
        B-CANF (ANFIC)           &   0.0 &  0.0 &&   0.0 &  0.0 &&   0.0 &  0.0\\
        Sheng'22 (ANFIC)         &  14.6 & 20.4 &&   9.2 & 13.6 &&  14.0 & 22.1\\
        Li'22 (ANFIC)            & -14.3 &  4.0 && -13.7 & -4.3 && -12.6 &  7.5\\
        \Xhline{2\arrayrulewidth}
    \end{tabular}
    \label{tab:BD-rate_I8}
    \color{black}
\end{table}

However, during testing, we employ a large GOP of size 32 (i.e. the intra-period is 32) with 5 hierarchical temporal levels in most cases. This results in a substantial number of B-frames being predicted from distant reference frames, leading to a significant domain shift (i.e. a significant change in statistics) between the training and test scenarios. This discrepancy can negatively affect the generalization of our ME-Net, motion prediction network, and frame synthesis network, especially when dealing with videos featuring substantial motion. In contrast, learned P-frame codecs like Li'22~\cite{Li_2022} and Sheng'22~\cite{tcm} have more consistent training and test scenarios in terms of temporal prediction distance. In P-frame coding, the domain shift can occur after coding several frames because the quality of the reference frame often degrades over time. In B-frame coding, it happens right from the first coding B-frame, affecting all the subsequent frames. This early domain shift in B-frame coding has a more immediate and pronounced impact on compression performance.

To mitigate the domain shift between the training and test scenarios, we present additional results in Table~\ref{tab:BD-rate_I8} for a smaller GOP size of 8. In order to ensure a fair comparison, we utilize ANFIC as the I-frame codec for all the competing methods. The results demonstrate that our B-CANF performs comparably to or better than Li'22~\cite{Li_2022} under GOP=8, even without utilizing advanced entropy coding. Note that these results are not intended to advocate the use of small GOPs or short intra-periods. Rather, they serve to highlight the limitations of Vimeo dataset~\cite{vimeo} in training B-frame codecs and demonstrate the potential of B-CANF.   

\color{black}

 \begin{table*}[t]
    \centering
    \setlength\tabcolsep{2.9pt}
    \caption{Ablation study of conditional coding. Setting A is our proposed B-CANF model.}
    \begin{tabular}{ccc@{}p{0.5em}@{}cc@{}p{0.5em}@{}cccc}
        \Xhline{2\arrayrulewidth}
    \multirow{2}{*}{Setting}    & \multicolumn{2}{c}{Motion coding} && \multicolumn{2}{c}{Inter-frame coding}  && \multicolumn{4}{c}{BD-rate (\%) PSNR-RGB} \\
         \cline{2-3} \cline{5-6} \cline{8-11}
         & Residual & Conditional && Residual & Conditional           && UVG     & MCL-JCV & HEVC-B  & CLIC'22 \\
         \Xhline{2\arrayrulewidth}
     A   &              & $\checkmark$ &&              & $\checkmark$ && -47.5   & -41.1   & -46.9   & -33.8 \\
     B   & $\checkmark$ &              &&              & $\checkmark$ && -41.8   & -36.3   & -42.2   & -30.0 \\
     C   &              & $\checkmark$ && $\checkmark$ &              && -33.4   & -26.8   & -34.6   & -23.0 \\
     D   & $\checkmark$ &              && $\checkmark$ &              && -28.2   & -22.8   & -30.3   & -20.0 \\
         \Xhline{2\arrayrulewidth}
    \end{tabular}
    \label{tab:cond_coding}
\end{table*}
\begin{table*}[t]
    \color{black}
    \centering
    \caption {Ablation study of the distortion weighting and the frame-type adaptation (FA) module. Setting A is our proposed B-CANF model.}

    \begin{tabular}{ccc@{}p{1em}@{}cccc}
        \Xhline{2\arrayrulewidth}
        \multirow{2}{*}{Setting} &
        \multirow{2}{*}{Distortion weighting} & \multirow{2}{*}{FA module}  && \multicolumn{4}{c}{BD-rate (\%) PSNR-RGB} \\
            \cline{5-8} &&&& UVG     & MCL-JCV & HEVC-B  & CLIC'22 \\
        \Xhline{2\arrayrulewidth}
        A & $\checkmark$ & $\checkmark$      && -47.5   & -41.1   & -46.9   & -33.8 \\
        B & $\checkmark$ &                   && -39.4   & -32.1   & -41.0   & -28.7 \\
        C &              & $\checkmark$      && -39.3   & -35.4   & -38.1   & -25.0 \\
        D &              &                   && -32.0   & -30.4   & -36.3   & -22.1 \\
        \Xhline{2\arrayrulewidth}
    \end{tabular}
    \label{tab:lambdaFA}
    \color{black}
\end{table*}
\subsection{Ablation Experiments} 
\label{sec:ablation}

We conduct ablation experiments with GOP 16 and intra-period 32. Unless otherwise specified, the BD-rates are reported against x265 (veryslow) in terms of PSNR-RGB.

\paragraph{Conditional vs. Residual Coding} Table~\ref{tab:cond_coding} analyzes the impact of our conditional coding scheme on compression performance. We test four variants, switching between residual and conditional coding for motion and inter-frame coding. When residual coding is selected, we use ANFIC~\cite{anfic} to code flow map or frame prediction residuals. Comparing variants A and D, turning both motion and inter-frame codecs into residual-type ones results in a considerable performance drop. In particular, our conditional coding scheme is seen to benefit inter-frame coding more than motion coding (from D to B vs. from D to C). A breakdown analysis of the bitstream composition explains that the motion part represents only 5-15\% of the entire bitstream. Interestingly, the gains of our conditional coding in improving motion coding and inter-frame coding are nearly additive. 

\begin{figure}[t!]
    \center
    \includegraphics[width=\linewidth]{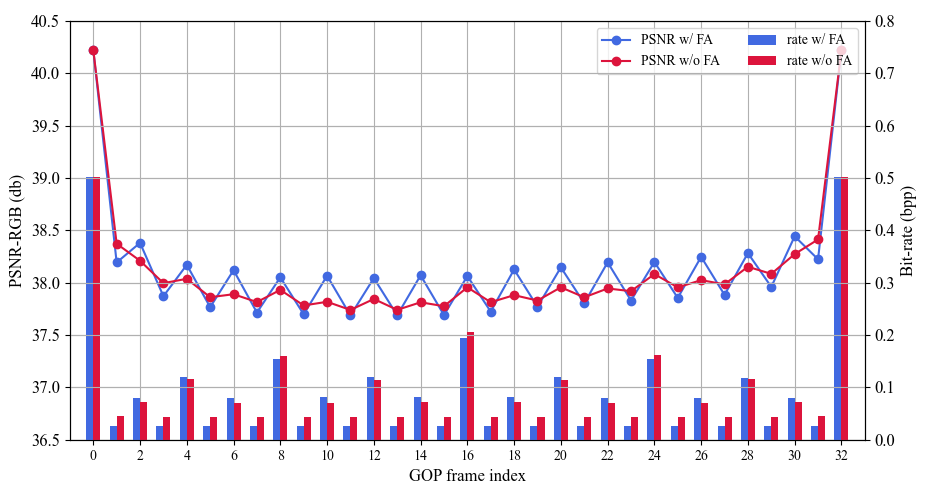}
    \caption{Profiles of the per-frame PSNR's and the per-frame bit rates averaged over GOPs on UVG dataset, with and without frame-type adaptive (FA) coding \textcolor{black}{at high rates ($\lambda_1=2048$). Frame 0, 32 are coded as I-frame, Frame 16 is coded as B*-frame, and the odd-numbered frames are non-reference B-frames while the even-numbered frames are reference B-frames.}}
    \label{fig:psnr_bitrate_gop_distribution}
\end{figure}

\paragraph{Frame-type Adaptive Coding} 
\label{sec:exp_abl_FA}
\textcolor{black}{As mentioned in Section~\ref{sec:FrameAdaptCoding}, there are two mechanisms that contribute to the better bit allocation of our B-CANF. The first is to weight differently the distortions of different types of B-frames during training. The second is the FA module, which allows both the conditional motion and inter-frame codecs to behave differently according to the
input frame’s type.}
\textcolor{black}{Table~\ref{tab:lambdaFA} presents an ablation study to shed lights on how they contribute to the rate-distortion performance. Comparing setting A (our B-CANF) with setting B, we see that turning off the FA module leads to a significant rate-distortion performance loss. In this case, the same (amortized/average) motion and inter-frame codecs are used to process different types of B-frames. Likewise, in comparison with the full model (setting A), having only the FA module without weighting the distortions according to the frame type (setting C) during training is sub-optimal too. These results suggest that they both contribute significantly to the resulting rate-distortion performance.}

Fig.~\ref{fig:psnr_bitrate_gop_distribution} \textcolor{black}{further} visualizes how \textcolor{black}{the proposed FA module} may impact the bit allocation among B-frames in a GOP and their decoded quality. \textcolor{black}{In this and the following experiments, the distortion weighting is enabled during training. We see that disabling the FA module (i.e., setting B in Table~\ref{tab:lambdaFA})} will learn an "average" codec that tends to allocate more bits to the non-reference B-frames. As a result, the PSNR distribution becomes more smooth and less hierarchical.

Table~\ref{tab:FA} further presents BD-rate results by turning on and off frame-type adaptive coding in the motion and/or inter-frame codecs. Interestingly, when applied to either the inter-frame codec (from D to C) or the motion codec (from D to B) alone, it is not as effective as the case when it is enabled for both codecs (from D to A).

\begin{figure}[t!]
    \center
    \includegraphics[width=\linewidth]{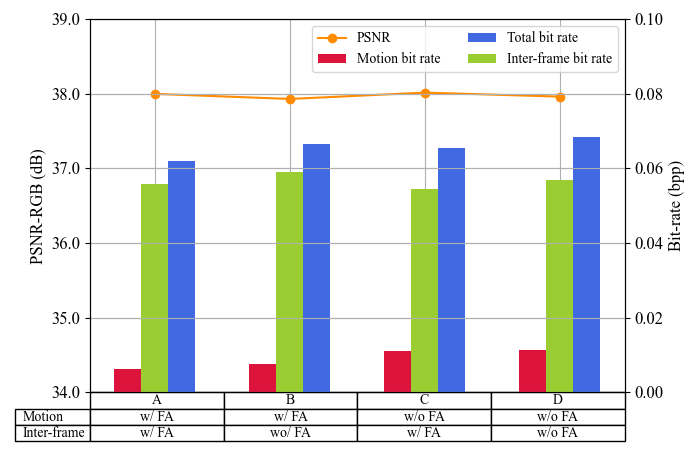}
    \caption{\textcolor{black}{Comparison of PSNR-RGB and bit allocation by turning on and off frame-type adaptive coding in the motion and/or inter-frame codecs at high rates ($\lambda_1=2048$). The PSNR-RGB and bit-allocation results are averaged over B-frames of the test sequences in UVG dataset. FA refers to frame-type adaptive coding. Setting A is our B-CANF model.}}
    \label{fig:fa_psnr_rate}
\end{figure}

\textcolor{black}{To gain insights into this observation, Fig.~\ref{fig:fa_psnr_rate} analyzes the bit allocation between motion coding and inter-frame coding for B-frames. Comparing setting C with setting D, we observe that enabling frame-type adaptive coding for the inter-frame codec only is able to reduce effectively the bit rate of inter-frame coding (green) while keeping the motion bit rate (red) relatively untouched. Likewise, the comparison of settings B and D shows that applying it to the motion codec only reduces effectively the motion bit rate (red). In this case, the slightly increased bit rate of inter-frame coding (green) may be attributed to the use of a shared inter-frame codec, which is unable to adapt to the changes in the inter-frame correlation that result from frame-type adaptive motion coding. Turing on frame-type adaptive coding for both codecs allows the greatest degree of flexibility in allocating bits to different parts, achieving a synergy effect.}

\begin{table*}[t]
    \centering
    \caption {\textcolor{black}{Ablation study of frame-type adaptive coding. Setting A is our proposed B-CANF model.}}

    \begin{tabular}{c@{}p{1em}cc@{}p{1em}@{}cccc}
        \Xhline{2\arrayrulewidth}
        \multirow{2}{*}{Setting} &&
        \multicolumn{2}{c}{Frame-type adaptive coding}  && \multicolumn{4}{c}{BD-rate (\%) PSNR-RGB} \\
            \cline{3-4} \cline{6-9}
          && Motion codec & Inter-frame codec && UVG     & MCL-JCV & HEVC-B  & CLIC'22 \\
        \Xhline{2\arrayrulewidth}
        A && $\checkmark$ & $\checkmark$      && -47.5   & -41.1   & -46.9   & -33.8 \\
        B && $\checkmark$ &                   && -41.6   & -32.7   & -40.6   & -28.4 \\
        C &&              & $\checkmark$      && -42.3   & -35.9   & -43.2   & -30.9 \\
        D &&              &                   && -39.4   & -32.1   & -41.0   & -28.7 \\
        \Xhline{2\arrayrulewidth}
    \end{tabular}
    \label{tab:FA}
\end{table*}
   
\begin{figure}[t]
    \begin{center}
        \subfloat[]{
            \centering
            \includegraphics[width=0.4\linewidth]{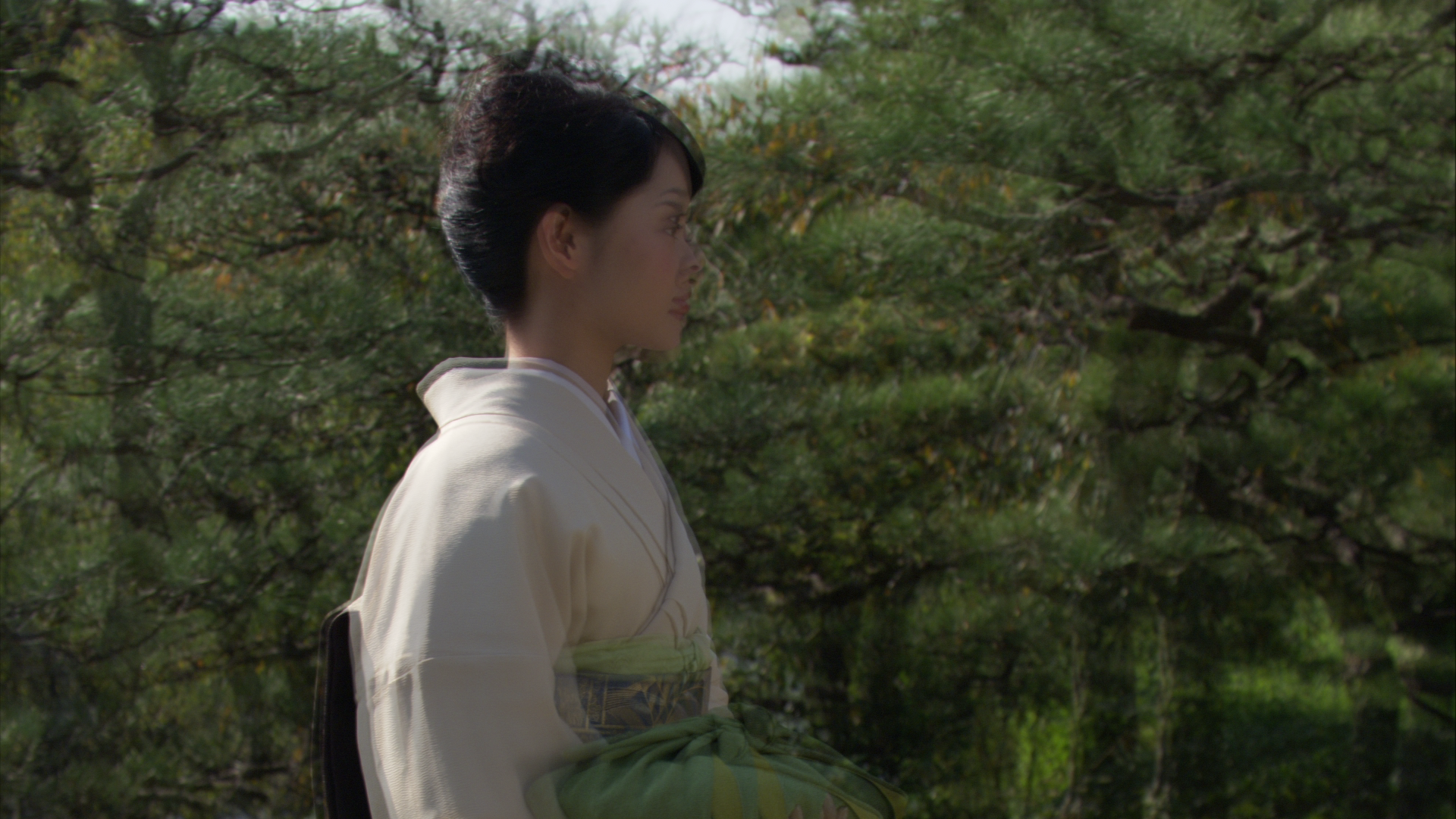} 
        }
        \subfloat[]{
            \centering
            \includegraphics[width=0.4\linewidth]{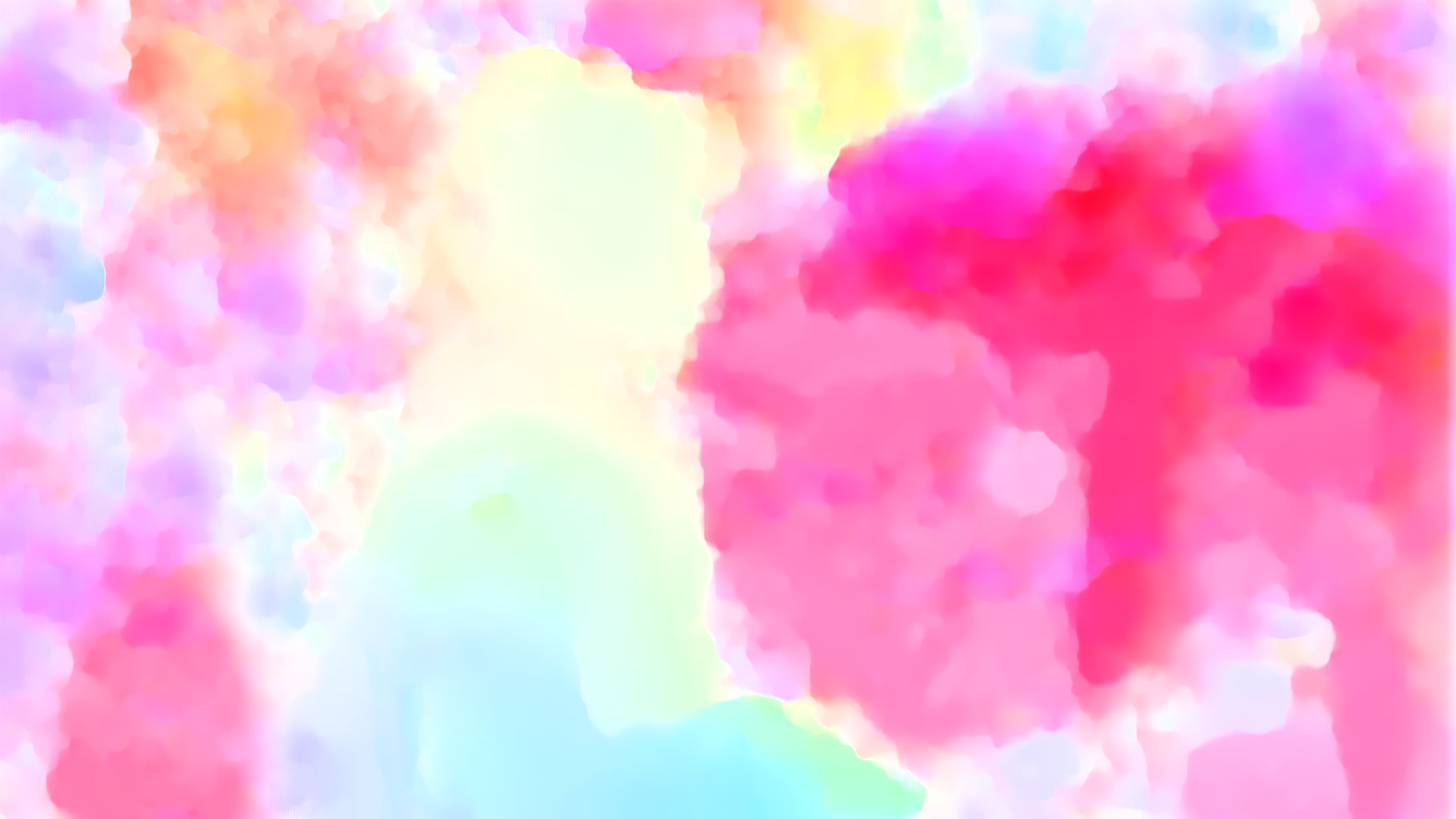} 
        }\\
        \subfloat[]{
            \centering
            \includegraphics[width=0.4\linewidth]{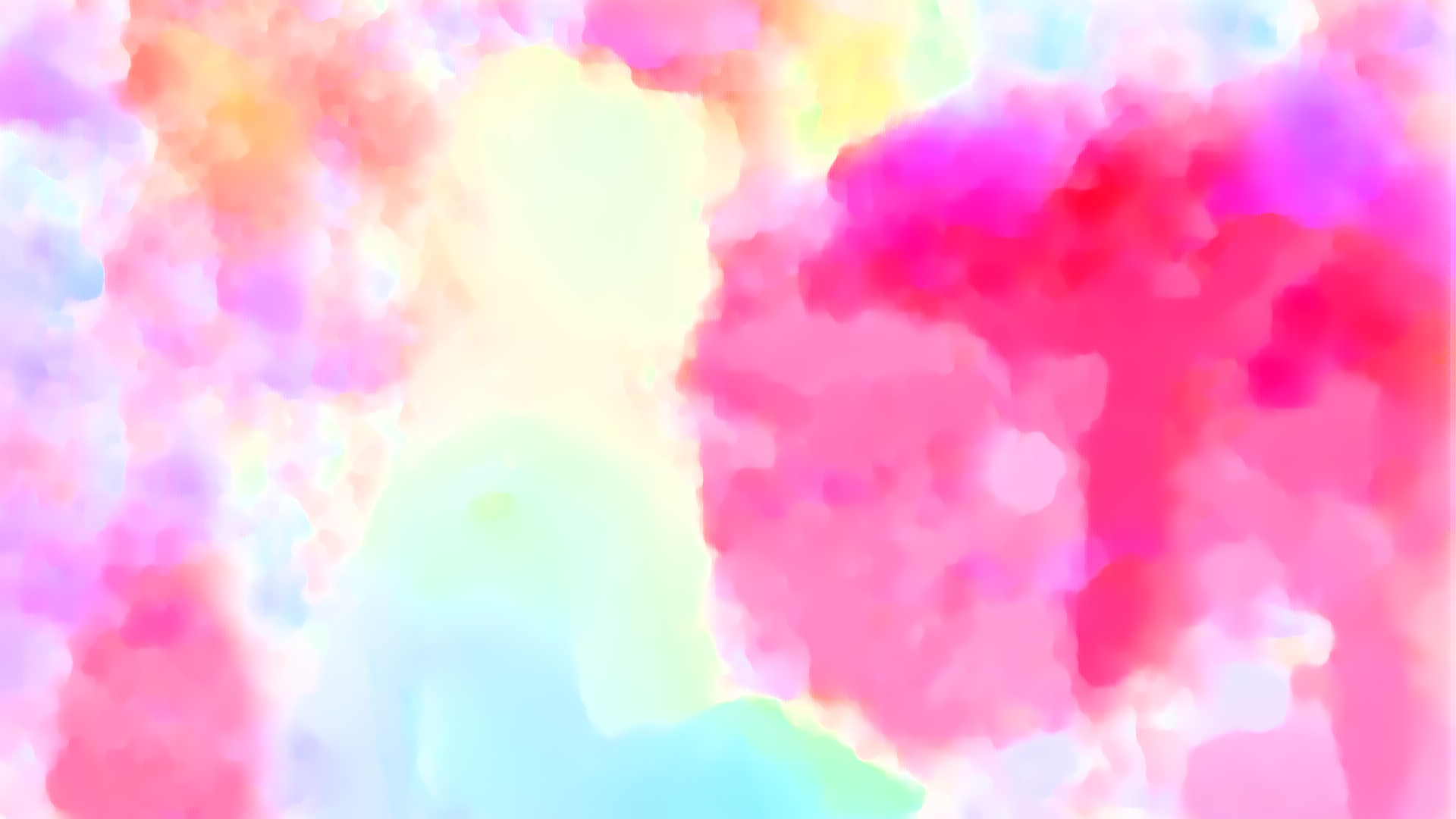}
        }
        \subfloat[]{
            \centering
            \includegraphics[width=0.4\linewidth]{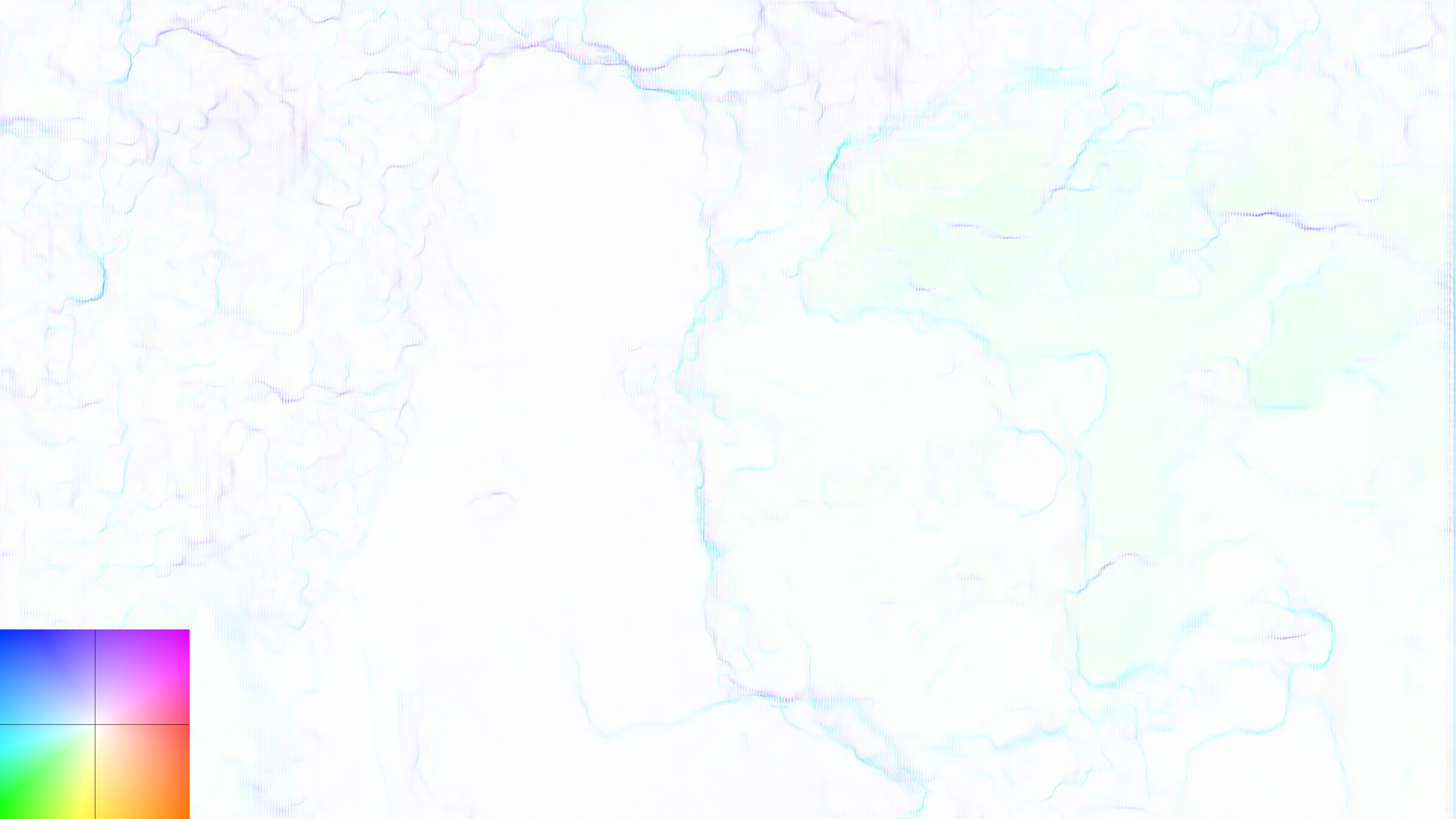}
        }
    \end{center}
    \caption{\textcolor{black}{Visualization of \textcolor{black}{(a) a coding frame overlaid with its reference frame,} (b) the coded flow map $\hat{m}^e_{t\to t-k}$ (cp.~Fig.~\ref{fig:overall}(b)), (c) its virtual counterpart after sign reversal, and (d) the differences between (b) and (c). White color represents zero motion.}} 
    \label{fig:ablation-pframe}
\end{figure}

\paragraph{Coded Flow Maps for B*-frames}   \label{sec:vis_B_frames}      Figs.~\ref{fig:ablation-pframe}\textcolor{black}{(b) and (c) visualize} the two coded flow maps for B*-frames. \textcolor{black}{As shown in Fig.~\ref{fig:ablation-pframe}(d), they} are seen to disagree slightly with each other around object boundaries. \textcolor{black}{Table~\ref{tab:abl_gb_flow} presents BD-rate results for three variants of motion compensation. \textit{Both} refers to using two hypotheses, i.e. the coded flow map $\hat{m}^e_{t\to t-k}$ and its virtual counterpart after sign reversal, for motion compensation, whereas \textit{First} or \textit{Second} refers to using solely one of them. We see that using solely $\hat{m}^e_{t\to t-k}$ (\textit{First}) or its virtual counterpart (\textit{Second}) for motion compensation leads to less rate saving than using both simultaneously. This may be attributed to the fact that the input flow map is usually less reliable in boundary regions. Performing multi-hypothesis prediction with slightly different motion estimates in these regions is an effective means of mitigating the reliability issue.}

\begin{table}[t]
    \caption{\textcolor{black}{Ablation study of the number of hypotheses used for motion compensating B*-frame under intra-period 32 and GOP size 4.}}
    \centering
    \tabcolsep=7pt
    \begin{tabular}{c@{}p{1em}@{}cccc}
        \Xhline{2\arrayrulewidth}
        \multirow{2}{*}{Hypothesis} && \multicolumn{4}{c}{BD-rate (\%) PSNR-RGB} \\
        \cline{3-6}
           && UVG     & MCL-JCV & HEVC-B  & CLIC'22 \\
        \Xhline{2\arrayrulewidth}
        Both             && -40.6 & -34.6 & -39.4 & -18.7 \\
        First            && -35.0 & -31.0 & -35.1 & -14.3 \\
        Second           && -35.2 & -30.9 & -35.1 & -14.7 \\
        \Xhline{2\arrayrulewidth}
    \end{tabular}
    \label{tab:abl_gb_flow}
\end{table}

\color{black}
\paragraph{B*-frame vs. Separate P-frame Coding}
\label{sec:B*vsP}
Table~\ref{tab:abl_gop} investigates the effectiveness of B*-frame coding as a substitute for P-frame coding under various GOP sizes. The former reuses the B-frame codec while the latter needs to train end-to-end a separate, dedicated P-frame codec together with the other components of B-CANF. In most cases, our B*-frame mechanism is seen to achieve similar BD-rate results to the use of a separate P-frame codec. In particular, under GOP size 4, it performs even better than P-frame coding. Recall that B*-frames are essentially a multi-hypothesis prediction technique.  

\begin{figure}
    \centering
    \includegraphics[width=0.8\linewidth]{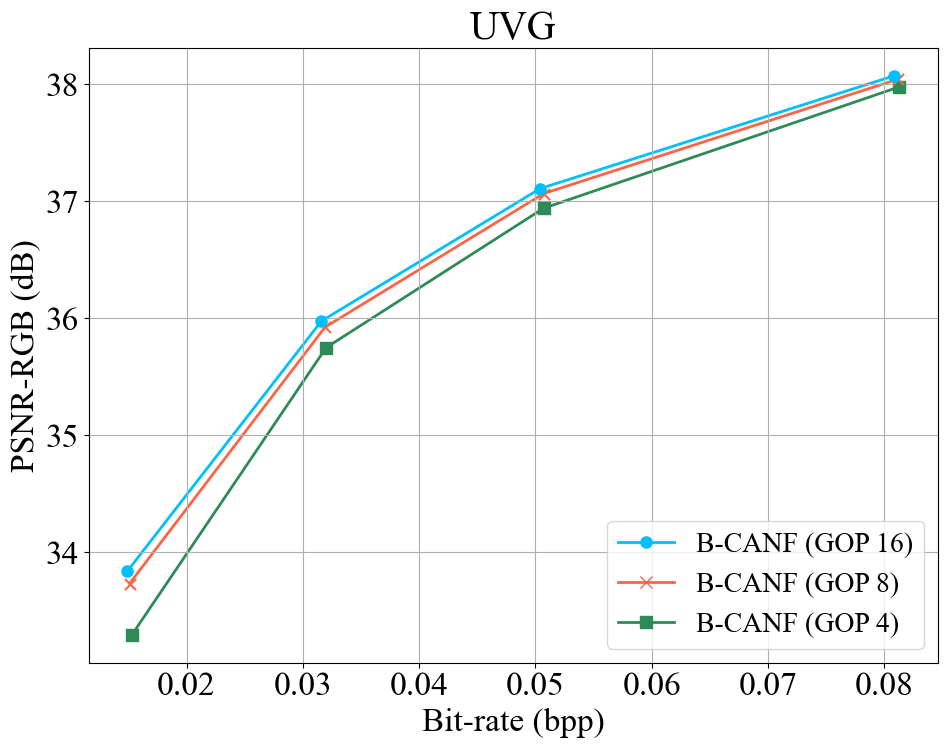}
    \caption{\textcolor{black}{Rate-distortion comparison of GOP sizes 4, 8, 16 on UVG dataset under intra-period 32.}}
    \label{fig:abl_gop_size}
\end{figure}

\paragraph{GOP Size vs. Rate-distortion Performance}
\label{sec:gopsize}
We evaluate the effect of GOP size on the performance of B-CANF. A number of GOP sizes, including 4, 8, 16, are tested with intra-period 32. The BD-rates are summarized in Table~\ref{tab:abl_gop} (see the results w/o a separate P-frame codec). The corresponding rate-distortion curves on UVG dataset are presented in Fig.~\ref{fig:abl_gop_size}. From Fig.~\ref{fig:abl_gop_size}, the rate-distortion performance of B-CANF is seen to improve with the increased GOP size. The improvement is most obvious at low rates. Like P-frames, our B*-frames suffer more from temporal error propagation with smaller GOP sizes (in which cases, B*-frames are sent more frequently), especially at low rates where poor reconstruction and motion quality is expected. Increasing GOP size decreases the frequency of B*-frames, thereby reducing temporal error propagation.
\begin{table}[t]
    \caption{\textcolor{black}{The effect of GOP size on B-CANF with and without a separate P-frame codec under intra-period 32.}}
    \centering
    \setlength{\tabcolsep}{5.2pt}
    \begin{tabular}{ccccccccc}
        \Xhline{2\arrayrulewidth}
        \multirow{3}{*}{GOP}    & \multicolumn{8}{c}{BD-rate (\%) PSNR-RGB} \\
        \cline{2-9}
                                &\multicolumn{2}{c}{UVG}     & \multicolumn{2}{c}{MCL-JCV} & \multicolumn{2}{c}{HEVC-B}  & \multicolumn{2}{c}{CLIC'22} \\
        \cline{2-9}
        & w/o & w/ & w/o & w/ & w/o & w/ & w/o & w/ \\
        \Xhline{2\arrayrulewidth}
        16      & \textbf{-47.5} & -46.4 & \textbf{-41.1} & \textbf{-41.1}  & -46.9 & \textbf{-47.7} & -33.8 & \textbf{-36.5} \\
        8       & \textbf{-45.7} & -44.8 & \textbf{-39.5} & -39.3 & -45.1 & \textbf{-46.3} & -29.7 & \textbf{-30.8} \\
        4       & \textbf{-40.6} & -34.7 & \textbf{-34.6} & -31.7 & \textbf{-39.4} & -37.2 & \textbf{-18.7} & -12.5 \\
        \Xhline{2\arrayrulewidth}
    \end{tabular}
    
    \label{tab:abl_gop}
\end{table}

\color{black}

\begin{table}[t]
    \color{black}
    \caption{Ablation study of the number of autoencoding trasforms}
    \centering
    \begin{tabular}{ccccc}
        \Xhline{2\arrayrulewidth}
        \multirow{2}{*}{\makecell[c]{Motion and \\ Inter-frame codecs}}            & \multicolumn{4}{c}{BD-rate (\%) PSNR-RGB} \\
        \cline{2-5}
                                        & UVG     & MCL-JCV & HEVC-B  & CLIC'22 \\
        \Xhline{2\arrayrulewidth}
        1-step B-CANF (CVAE) & -36.7 & -30.1 & -37.3 & -24.9 \\
        2-step B-CANF (Ours) & -47.5 & -41.1 & -46.9 & -33.8 \\
        \Xhline{2\arrayrulewidth}
    \end{tabular}
    \color{black}
    \label{tab:abl_numAuto}
\end{table}
\begin{figure*}[t!]
\centering
\resizebox{0.845\textwidth}{!}
{
\LARGE
\begin{tabular}{ccccc}
    Ground Truth & HM & LHBDC (MSE) & B-CANF (MSE) & B-CANF (SSIM) \\

    \includegraphics[width=0.3\textwidth]{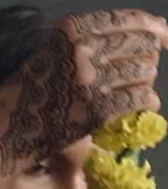}&
    \includegraphics[width=0.3\textwidth]{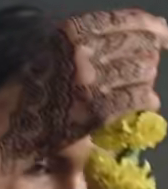}&
    \includegraphics[width=0.3\textwidth]{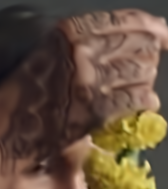}&
    \includegraphics[width=0.3\textwidth]{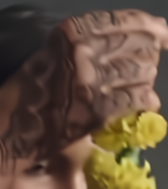}&
    \includegraphics[width=0.3\textwidth]{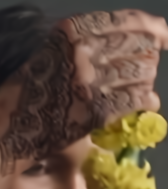}\\
     & PSNR: $37.72$ dB & PSNR: $37.29$ dB & PSNR: $37.41$ dB & \\
     & MS-SSIM: $0.9744$ dB & & & MS-SSIM: $0.9833$ dB \\
     & \textcolor{black}{$0.0025$} bpp & \textcolor{black}{$0.0146$} bpp & \textcolor{black}{$0.0025$} bpp & \textcolor{black}{$0.0016$} bpp \\
    \includegraphics[width=0.3\textwidth]{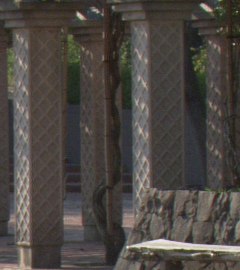}&
    \includegraphics[width=0.3\textwidth]{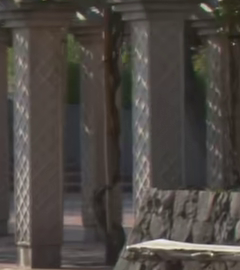}&
    \includegraphics[width=0.3\textwidth]{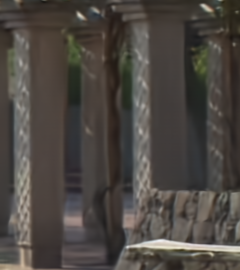}&
    \includegraphics[width=0.3\textwidth]{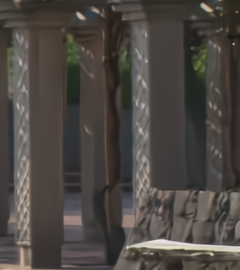}&
    \includegraphics[width=0.3\textwidth]{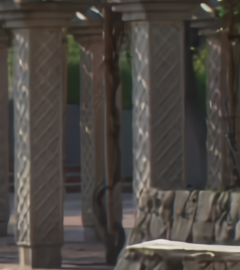}\\
     & PSNR: $31.72$ dB & PSNR: $31.21$ dB & PSNR: $31.47$ dB & \\
     & MS-SSIM: $0.9362$ dB & & & MS-SSIM: $0.9437$ dB \\
     & \textcolor{black}{$0.0027$} bpp & \textcolor{black}{$0.0255$} bpp & \textcolor{black}{$0.0029$} bpp & \textcolor{black}{$0.0015$} bpp \\
    \includegraphics[width=0.3\textwidth]{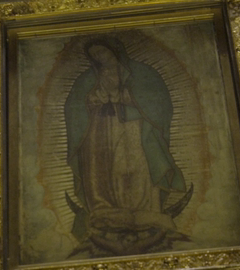}&
    \includegraphics[width=0.3\textwidth]{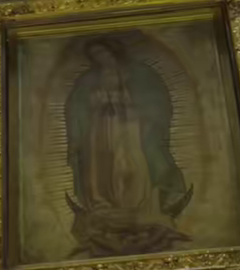}&
    \includegraphics[width=0.3\textwidth]{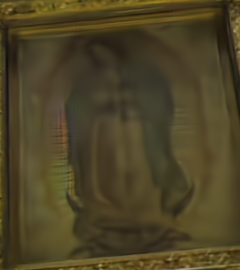}&
    \includegraphics[width=0.3\textwidth]{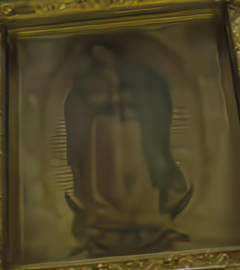}&
    \includegraphics[width=0.3\textwidth]{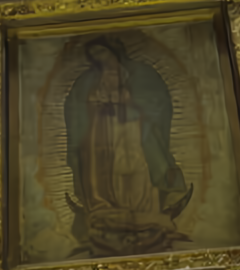}\\
     & PSNR: $36.23$ dB & PSNR: $34.42$ dB & PSNR: $35.60$ dB & \\
     & MS-SSIM: $0.9668$ dB & & & MS-SSIM: $0.9784$ dB \\
     & \textcolor{black}{$0.0016$} bpp & \textcolor{black}{$0.0282$} bpp & \textcolor{black}{$0.0023$} bpp & \textcolor{black}{$0.0018$} bpp \\
\end{tabular}
}

\caption{Subjective quality comparison of HM (randomaccess), LHBDC~\cite{murat_lhbdc} and our B-CANF. The parentheses "(MSE)" and "(SSIM)" indicate that the model is trained with MSE and MS-SSIM, respectively. Zoom in for better visualization.}

\label{fig:visual}
\end{figure*}
\begin{table*}[t!]
    \caption{\textcolor{black}{Comparison of the model complexity in terms of encoding/decoding  MACs, runtimes, model size, decoded picture buffer, and peak memory. "full-res" stands for the size of input resolution.}}
    \centering
    \begin{tabular}{cccc@{}p{0.5em}ccccc}
        \Xhline{2\arrayrulewidth}
        \multirow{2}{*}{Method} & \multirow{2}{*}{Frame-type} & \multicolumn{2}{c}{Encode} && \multicolumn{2}{c}{Decode} & Model & Decoded  & Peak \\
        \cline{3-4} \cline{6-7}
               &      & Time  & MACs && Time  & MACs & Size & Picture Buffer & Memory  \\
        \Xhline{2\arrayrulewidth}
        HM~\cite{HM}        &  B    & 33.28s & - && 0.04s  & - & - & - & -\\
        VTM~\cite{vtm}      &  B    & 731.33s & - && 0.07s  & - & - & - & -\\
        \Xhline{2\arrayrulewidth}
        DCVC~\cite{dcvc}            &  P    & 7.70s & 1.16M/pixel && 28.97s & 0.77M/pixel & 8M  & 3 full-res & 128 full-res\\
        CANF-VC~\cite{CANF-VC}      &  P    & 1.45s & 2.45M/pixel && 1.07s  & 1.77M/pixel & 31M & 13 full-res & 64 full-res\\
        
        Sheng'22~\cite{tcm}         &  P    & 0.82s & 1.42M/pixel && 0.58s  & 0.92M/pixel & 10.7M  & 67 full-res & 128 full-res\\
        Li'22~\cite{Li_2022}        &  P    & 0.83s & 1.68M/pixel && 0.61s  & 1.25M/pixel & 17.5M  & 67 full-res & 128 full-res\\
        
        LHBDC~\cite{murat_lhbdc}    &  B    & 1.19s & 1.70M/pixel && 0.73s  & 1.12M/pixel & 23.5M  & 15 full-res &96 full-res\\
        B-CANF (Ours)               &  B*   & 1.44s & 2.68M/pixel && 1.06s  & 2.01M/pixel & 24M    & 3 full-res & 64 full-res\\
        B-CANF (Ours)               &  B    & 1.69s & 3.08M/pixel && 1.09s  & 2.10M/pixel & 24M    & 15 full-res  & 64 full-res\\
        \Xhline{2\arrayrulewidth}
    \end{tabular}
    \label{tab:exp_comlexity}
\end{table*}

\color{black}
\paragraph{The Number of Autoencoding Transforms}
\label{sec:numberAuto}
Table~\ref{tab:abl_numAuto} compares the BD-rates between 2-step and 1-step B-CANF to explore the effect of the number of autoencoding transforms on compression performance. The 1-step B-CANF is obtained by skipping the first autoencoding transform \{$A_1, S_1$\} in Fig.~\ref{fig:anfs}(b). In order to have a similar model size to the 2-step B-CANF, it has more channels in each autoencoding transform. Remarkably, the 1-step B-CANF reduces to a specific implementation of conditional VAE (CVAE). From Table~\ref{tab:abl_numAuto}, we observe that the gain of the 2-step B-CANF over the 1-step B-CANF is obvious across the datasets. The result justifies our use of CANF rather than CVAE.
\color{black}
\subsection{Subjective Quality}

\color{black}
Fig.~\ref{fig:visual} presents the subjective quality comparison between HM (randomaccess), LHBDC~\cite{murat_lhbdc} and our B-CANF. LHBDC (MSE) and B-CANF (MSE) are trained to optimize PSNR-RGB while B-CANF (SSIM) is our model trained with MS-SSIM-RGB. All the schemes are evaluated with GOP 16 and intra-period 32, and all the learning-based methods use ANFIC as the I-frame codec. It is seen that our B-CANF (MSE) achieves comparable or even better subjective quality than LHBDC (MSE), with its bit rate being nearly one order of magnitude smaller than that of LHBDC (MSE). Compared with HM (randomaccess), our B-CANF (SSIM) preserves more texture details 
(cf. patterns on fingers in the first row, pillars in the second row, and textures in the last row) at a lower bit rate.
\color{black}

\subsection{Complexity Analysis}
\color{black}
Table~\ref{tab:exp_comlexity} characterizes the complexity of the proposed B-CANF in terms of runtimes, MACs and model size. We also measure the memory footprint following~\cite{qual_ttcoding}. This includes the size of the decoded picture buffer, which stores decoded frames, flow maps, and/or context features. The unit of measurement is "full-res," where one reconstructed frame occupies the equivalent of 3 full-res. The peak memory represents the maximum buffering requirement for generated features during compression.
The reported numbers in Table~\ref{tab:exp_comlexity} were obtained using an NVIDIA GeForce RTX 2080Ti GPU for the learned codecs, and an Intel(R) Core(TM) i7-9700K CPU @ 3.60GHz for HM~\cite{HM} and VTM~\cite{vtm}. The experiments were conducted using 1080p input videos.

For frame-type P/B*, we use the prediction structure IPPP... under intra-period 32 and GOP size 32. The encoding/decoding times are then averaged over the first 100 P/B*-frames of Beauty sequence in UVG dataset. For frame-type B, the encoding/decoding times are averaged over the first 100 B-frames of the same sequence with intra-period 32, GOP size 16, and hierarchical B prediction. 

\color{black}
The following observations can be made from Table~\ref{tab:exp_comlexity}. (1) The prolonged runtimes of DCVC~\cite{dcvc} are attributed to the use of an autoregressive model for entropy coding. In contrast, CANF-VC~\cite{CANF-VC}, LHBDC~\cite{murat_lhbdc} and our B-CANF do not use any auto-regressive model for motion coding or inter-frame coding. (2) In comparison with VAE-based coding, DCVC~\cite{dcvc} and LHBDC~\cite{murat_lhbdc}, the larger MAC of CANF-VC~\cite{CANF-VC} and B-CANF comes from stacking multiple autoencoding transforms. Nevertheless, the comparable encoding/decoding runtimes suggest that these autoencoding transforms are parallel-friendly. (3) Our B-CANF has lower buffering requirements and peak memory when compared to other approaches.  (4)
With our B-CANF, the encoding time and MAC of B-frames are seen to be higher than those of B*-frames (the last two rows in~Table~\ref{tab:exp_comlexity}). This is because motion estimation is performed twice for B-frames, whereas it is carried out only once for B*-frames (just like P-frames). In contrast, the decoding times and MAC of both types of frame are quite close, even though B-frames incur extra computations in the motion prediction network (Fig.~\ref{fig:overall}(a) vs. Fig.~\ref{fig:overall}(b)). This indicates that the motion prediction network is not the major computation bottleneck during decoding. %Note that the prolonged runtimes of DCVC~\cite{dcvc} are attributed to the use of an autoregressive model for entropy coding.
(5) As compared with CANF-VC~\cite{CANF-VC}, which supports only P-frame coding, the B-frame coding in B-CANF has higher encoding MAC because motion estimation is performed twice. In contrast, our B*-frames, similar to P-frames, have comparable encoding MAC to CANF-VC~\cite{CANF-VC}. The decoding MAC's of both B-frames and B*-frames are seen to be comparable to that of CANF-VC~\cite{CANF-VC}. (6) In terms of model size, B-CANF is similar to LHBDC~\cite{murat_lhbdc}.

\color{black}

\section{Conclusion}
\label{sec:conclusion}

\textcolor{black}{In this paper, we propose a CANF-based B-frame coding framework, known as B-CANF, that exploits the notion of conditional coding for both motion and inter-frame coding. It features B*-frames and frame-type adaptive coding. We show that (1) B*-frames allow greater flexibility in supporting various GOP sizes without the need for an extra P-frame codec, and (2) frame-type adaptive coding improves the bit allocation among B/B*-frames. Extensive experimental results confirm the superiority of B-CANF to the other state-of-the-art B-frame coding schemes. How to achieve even higher compression performance by addressing the domain shift issue and reducing the model's complexity is our future work.}

\bibliography{egbib}
\bibliographystyle{IEEEtran}

\begin{IEEEbiography}[{\includegraphics[width=1in,height=1.25in,clip,keepaspectratio]{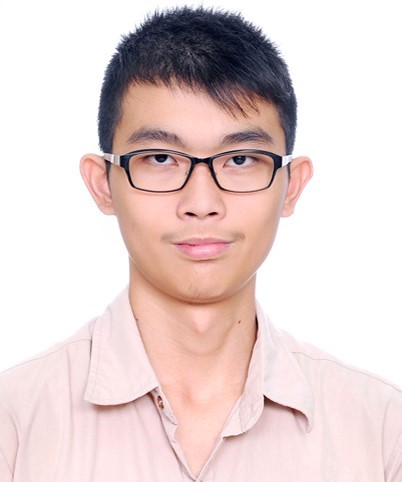}}]{Mu-Jung Chen} 
received his B.S. degree in electronic and computer engineering from National Taiwan University of Science and Technology (NTUST), Taiwan, in 2021. He is currently pursuing his M.S. degree in computer science and engineering, National Yang Ming Chiao Tung University (NYCU). His research interests include learning-based image/video coding, computer vision, and deep/machine learning.
\end{IEEEbiography}

\vspace{-3em}

\begin{IEEEbiography}[{\includegraphics[width=1in,height=1.25in,clip,keepaspectratio]{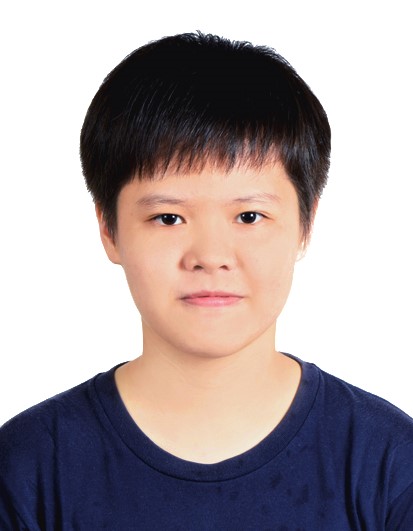}}]{Yi-Hsin Chen} 
received her B.S. degree in applied mathematics from National Chung Hsing University (NCHU), Taiwan, and her M.S. degree in data science and engineering from National Chiao Tung University (NCTU), Taiwan, in 2018 and 2020, respectively. She is currently pursuing her Ph.D. degree in computer science and engineering, National Yang Ming Chiao Tung University (NYCU), Taiwan. Her research interests include learning-based image/video coding, image/video restoration, computer vision, and deep/machine learning.
\end{IEEEbiography}
\vspace{-3em}

\begin{IEEEbiography}[{\includegraphics[width=1in,height=1.25in,clip,keepaspectratio]{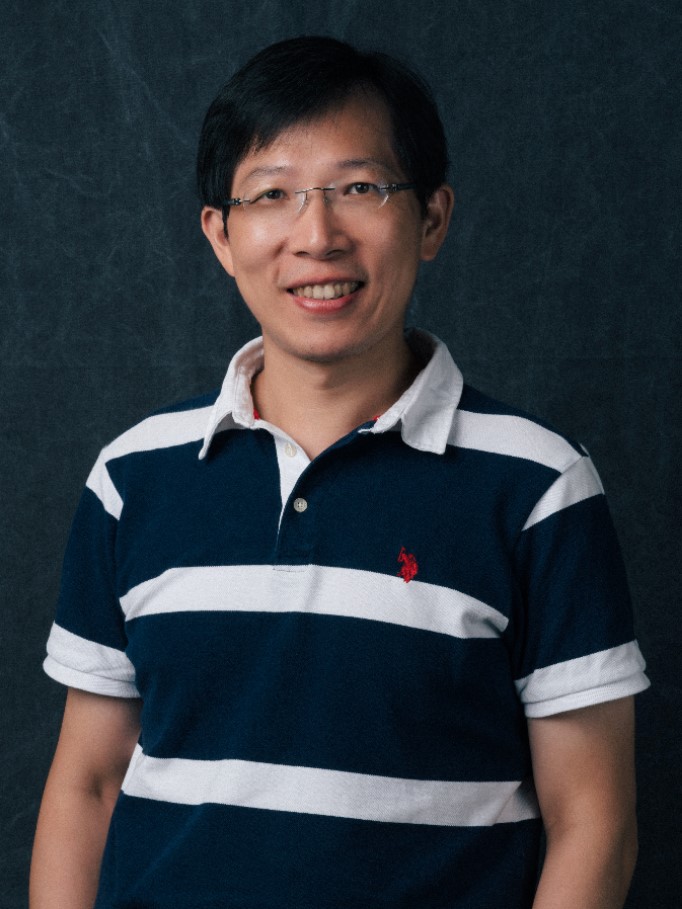}}]{Wen-Hsiao Peng} 
received his Ph.D. degree from National Chiao Tung University (NCTU), Taiwan, in 2005. He was with the Intel Microprocessor Research Laboratory, USA, from 2000 to 2001, where he was involved in the development of ISO/IEC MPEG-4 fine granularity scalability. Since 2003, he has actively participated in the ISO/IEC and ITU-T video coding standardization process and contributed to the development of SVC, HEVC, and SCC standards. He was a Visiting Scholar with the IBM Thomas J. Watson Research Center, USA, from 2015 to 2016. He is currently a Professor with the Computer Science Department, National Yang Ming Chiao Tung University, Taiwan. He has authored over 75+ journal/conference papers and over 60 ISO/IEC and ITU-T standards contributions. His research interests include learning-based video/image compression, deep/machine learning, multimedia analytics, and computer vision. Dr. Peng was Chair of the IEEE Circuits and Systems Society (CASS) Visual Signal Processing (VSPC) Technical Committee from 2020-2022. He was Technical Program Co-chair for 2021 IEEE VCIP, 2011 IEEE VCIP, 2017 IEEE ISPACS, and 2018 APSIPA ASC; Publication Chair for 2019 IEEE ICIP; Area Chair/Session Chair/Tutorial Speaker/Special Session Organizer for IEEE ICME, IEEE VCIP, and APSIPA ASC; and Track/Session Chair and Review Committee Member for IEEE ISCAS. He served as AEiC for Digital Communications for IEEE JETCAS and Associate Editor for IEEE TCSVT. He was Lead Guest Editor, Guest Editor and SEB Member for IEEE JETCAS, and Guest Editor for IEEE TCAS-II. He was Distinguished Lecturer of APSIPA and the IEEE CASS.
\end{IEEEbiography}

\vfill
\end{document}